\documentclass{article}%
\usepackage[dvips]{graphicx}
\usepackage{amsmath}%
\usepackage{amsfonts}%
\usepackage{amssymb}

\begin{document}

\begin{center}
{\Large Mechanism of antifreeze proteins action, based on}

{\Large Hierarchic theory of water and new ''clusterphilic''
interaction\smallskip}

\medskip

{\large Alex Kaivarainen\medskip}

\smallskip

\textbf{H2o@karelia.ru}

\textbf{\thinspace http://www.karelia.ru/\symbol{126}alexk}

\bigskip
\end{center}

\ 

\textbf{ }A basically new Hierarchic theory, general for solids and liquids
(Kaivarainen, 2001, 2000, 1995, 1992), has been briefly described and
illustrated by computer simulations on examples of water and ice. Full
description of theory and its numerous applications are presented in series of
articles at the arXiv of Los-Alamos (see \ http://arXiv.org/abs/physics/01022086).

\textbf{New clusterphilic interactions}, intermediate between hydrophilic and
hydrphobic, are introduced. They can be subdivided into:
\textbf{intramolecular} - when water cluster is localized in the ''open''
states of big interdomain or intersubunit cavities and\textbf{ intermolecular}
clusterphilic interactions. Intermolecular clusterphilic interactions can be
induced by very different macromolecules. The latter displays themselves in
bordering of water cluster by macromolecules and forming so-called ''clustrons''.

Clusterphilic interactions can play an important role in self-organization of
biosystems, especially multiglobular allosteric enzymes, microtubules and the
actin filaments. Cooperative properties of the cytoplasm, formation of
thixotropic structures, signal transmission in biopolymers, membranes and
distant interactions between different macromolecules may be mediated by both
types of clusterphilic interactions.

The selected review of literature, devoted to antifreeze proteins (AFP) and
ice-nucleation proteins (INP) interaction with water is presented.
Corresponding experimental results were analyzed on the base of Hierarchic
theory. The possible mechanism of cryoproteins influence on water, changing
its freezing point has been proposed. The consequences of new model of AFP
action and ways of its experimental verification are described also.\bigskip

\begin{center}
\medskip\medskip\textbf{I. }\ \textbf{INTRODUCTION }
\end{center}

A quantum and quantitative theory of liquid state, as well as a general theory
of condensed matter, was absent till now. This fundamental problem is crucial
for different brunches of science and technology. The existing solid states
theories did not allow to extrapolate them successfully to liquids.

Widely used molecular dynamics method is based on classical approach and
corresponding computer simulations. It cannot be considered as a general one.
The understanding of hierarchic organization of matter and developing of
general theory include a mesoscopic bridge between microscopic and macroscopic
properties of condensed matter. The biggest part of molecules of solids and
liquids did not follow classical Maxwell-Boltzmann distribution. This means
that only quantum approach is valid for elaboration of general theory of
condensed matter.

Our theoretical study of \textbf{water and aqueous systems }was initiated in
1986. It was stimulated by necessity to explain the nontrivial phenomena,
obtained by different physical methods in our investigations of water-protein
solutions. For example, the temperature anomalies in water physical
properties, correlating with changes in large scale protein dynamics were
found in our group by specially elaborated experimental approaches
(Kaivarainen, 1985). It becomes evident, that the water clusters and water
hierarchical cooperative properties are dominating factors in
self-organization, function and evolution of biosystems. The living organisms
are strongly dependent on water properties, representing about 70\% of the
body mass.

Due to its numerous anomalies, water is an ideal system for testing a new
theory of condensed matter. If the theory works well with respect to water and
ice, it is very probable, that it is valid for other liquids, glasses or
crystals as well. For this reason we have made the quantitative verification
of our hierarchic concept (Kaivarainen, 1989, 1992, 1995, 1996, 2000,
2001)\ on examples of water and ice.

Our theory considers two main types of molecular heat motion:
\textit{translational (tr) }and \textit{librational (lb) anharmonic
}oscillations, which are characterized by certain distributions\textbf{\ }in
three- dimensional (3D) impulse space. The most probable impulse or momentum
(p) determine the \textit{most probable }de Broglie wave (wave B) length
$(\lambda_{B}=h/p=v_{ph}/\nu_{B})$ and phase velocity $(v_{ph})$.
Conformational intramolecular dynamics is taken into account indirectly, as
far it has an influence on the intermolecular dynamics and parameters of waves
B in condensed matter. Solids and liquids are considered as a hierarchical
system of collective excitations - metastable quasiparticles of the four new
type: \textbf{effectons, transitons, convertons and deformons,}\textit{\ }%
strongly interrelated with each other.

When the length of standing waves B of molecules exceed the distances between
them, then the coherent molecular clusters may appear as a result of high
temperature molecular Bose-condensation (BC). The possibility of BC in liquids
and solids at the ambient temperatures is one of the most important results of
our model, confirmed by computer simulations. Such BC is mesoscopic one, in
contrast to macroscopic BC, responsible for superfluidity and superconductivity.

The interaction between atoms and molecules in condensed matter is much
stronger and thermal mobility/impulse much lesser, than in gas phase. It means
that the temperature of Bose condensation can be much higher in solids and
liquids than in the gas phase. \textbf{The lesser is interaction between
molecules or atoms the lower temperature is necessary for initiation of Bose condensation.}

This is confirmed in 1995 by Ketterle's group in MIT and later by few other
groups, showing the Bose-Einstein condensation in gas of neutral atoms, like
sodium (MIT), rubidium (JILA) and lithium (Rice University) at very low
temperatures, less than 1$^{0}$K. However, at this temperatures the number of
atoms in the primary effectons (Bose condensate) was about 20,000 and the
dimensions were almost macroscopic: about 15 micrometers.

For comparison, the number of water molecules in primary librational effecton
(coherent cluster), resulting from mesoscopic BC at freezing point 273 $^{0}$K
is only 280 and the edge length about 20 \AA\ (see Fig. 7).

\smallskip

Our \textbf{Hierarchic theory of matter }unites and extends strongly two
earlier existing most general models of solid state (Ashkroft and Mermin, 1976):

a) the Einstein model of condensed matter as a system of independent quantum oscillators;

b) the Debye model, taking into account only collective phenomena - phonons in
a solid body as in continuous medium.

Among earlier models of liquid state the model of flickering clusters by Frank
and Wen (1957) is closest of all to our model. In our days the quantum field
theoretical approach to description of biosystems with some ideas, close to
our ones has been developed intensively by Umesawa's group (Umezawa, et. al.,
1982; Umezawa, 1993) and Italian group (Del Giudice, et al., 1983; 1988, 1989).

Arani et al. introduced in 1998 the notion of Coherence Domains (CD), where
molecules are orchestrated by the internal electromagnetic waves (IR photons)
of matter. This idea is close to our notion of collective excitations of
condensed matter in the volumes of 3D translational and librational IR
photons, named primary electromagnetic deformons (see next section).\smallskip

\textbf{The new physical ideas require a new terminology}. It is a reason why
one can feel certain discomfort at the beginning of this work reading. To
facilitate this process, we present below a description of a new
quasiparticles, notions and terminology, introduced in our Hierarchic Theory
of matter (see Table 1). Most of notions and excitation properties, presented
below are not postulated,\textit{\ }but are the result of our computer
simulations.\medskip\ 

\begin{center}
\ \textbf{\ \ II.}\ \ \textbf{THE NEW NOTIONS AND DEFINITIONS, INTRODUCED IN }

\textbf{HIERARCHIC THEORY OF MATTER }\smallskip
\end{center}

\textbf{The most probable de Broglie wave (wave B). }

In composition of condensed matter the dynamics of particles could be
characterized by the thermal anharmonic oscillations of two types:
\textit{translational }(tr) and \textit{librational }(lb).

The length of the most probable wave B of thermally activated molecule, atom
or atoms group in condensed matter can be estimated by two following ways:%

\begin{equation}
(\lambda_{1,2,3}=h/mv_{gr}^{1,2,3}=v_{ph}^{1,2,3}/\nu_{B}^{1,2,3})_{tr,lb}
\tag{2.1}%
\end{equation}
where the most probable impulse $p^{1,2,3}=mv_{gr}^{1,2,3}$ is equal to
product of the particle mass (m) and most probable group velocity
$(v_{gr}^{1,2,3})$. The wave B length also could be evaluated as the ratio of
its most probable phase velocity $(v_{ph}^{1,2,3})$ to most probable frequency
$(\nu_{B}^{1,2,3})$.

The indices (1,2,3) correspond to selected directions of motion in 3D space,
related to the main axes of the molecules symmetry and their tensor of
polarizability. In the case when molecular motion is anisotropic one, we have inequalities:%

\begin{equation}
\lambda_{B}^{1}\neq\lambda_{B}^{2}\neq\lambda_{B}^{3} \tag{2.2}%
\end{equation}
Due to anharmonicity of oscillations: $(mv^{2}/2)<(kT/2)$ - the most probable
kinetic energy of molecules $(T_{\text{kin}})_{tr,lb}$ is lesser than
potential one $(V)_{tr,lb}$. Consequently, the most probable wave B length may
be bigger than space, occupied by one molecule in condensed matter: \
\begin{equation}
(V_{0}/N_{0})^{1/3}<\,\lambda_{B}^{1,2,3}>h/(mk_{B}T)^{1/2} \tag{2.2a}%
\end{equation}
where $V_{0}$ and $N_{0}$ are molar volume and Avogadro number
correspondingly. The left part of this inequality represents condition of
mesoscopic Bose condensation.

\smallskip

\textbf{\ The most probable (primary) effectons (tr and lb). }Such a new type
of quasiparticles (excitations) is represented by 3D superposition of three
most probable standing waves B of molecules. The shape of primary effectons in
a general case can be approximated by a parallelepiped, with the length of its
3 edges determined by 3 most probable waves B length. The volume of primary
effectons is equal to:%

\begin{equation}
V_{ef}=(9/4\pi)\lambda_{1}\lambda_{2}\lambda_{3}. \tag{2.3}%
\end{equation}
The number of molecules or atoms forming effectons is:%

\begin{equation}
n_{m}=(V_{ef})/(V_{0}/N_{0}), \tag{2.4}%
\end{equation}
where V$_{0}$ and $N_{0}$ are molar volume and Avogadro number,
correspondingly. The $n_{m}$ increases with temperature decreasing and may
reach hundreds or even thousands of molecules.

In liquids, primary effectons may be checked as a clusters and in solids as
domains or microcrystalline.

The thermal oscillations in the volume of corresponding effectons are
synchronized. It means the coherence of the most probable wave B of molecules
and their wave functions. We consider the \textbf{primary effectons as a
result of partial Bose condensation }of molecules of condensed matter. Primary
effectons correspond to the main state of Bose-condensate with the packing
number $n_{p}= 0$, i.e. with the resulting impulse equal to zero.

\textit{Primary effectons, as a coherent clusters, represent self-organization
of condensed matter on mesoscopic level, like I. Prigogin dissipative
structures. However, it is a quantum phenomenon. }

\textit{The volume of primary translational effectons in liquids is less than
the volume, occupied by one molecule }$(V_{0}/N_{0})$\textit{, this points to
the classical behavior of molecules.}

\textbf{''Acoustic'' (a) and ''optical'' (b) states of primary effectons. }

The ''acoustic'' \textit{a}-state of the effectons is such a dynamic state
when molecules or other particles composing the effectons, oscillate in the
same phase (i.e. without changing the distance between them).

The ''optic'' \textit{b}-state of the effectons is such dynamic state when
particles oscillate in the counterphase manner (i.e. with periodical change of
the distance between particles). This state of primary effectons has a common
features with Fr\"{o}lich's mode (Fr\"{o}lich, 1988).

The kinetic energies of ''acoustic'' (\textit{a}) and ''optical'' (\textit{b})
states are equal $[T_{\text{kin}}^{a}=T_{\text{kin}}^{b}]$ in contrast to
potential energies $[V^{a}<V^{b}]$. It means that the most probable impulses
in (\textit{a}) and (\textit{b}) states and, consequently, the wave B length
and \textbf{spatial dimensions of the effectons in the both states are equal.}
The energy of intermolecular interaction (Van der Waals, Coulomb, hydrogen
bonds etc.) in \textit{a}-state are bigger than that in \textit{b}-state.
Consequently, the molecular polarizability in \textit{a}-state also could be
bigger than in \textit{b}-state. It means that dielectric properties of matter
may change as a result of shift of the $(a\Leftrightarrow b)_{tr,lb}$
equilibrium of the effectons.

\textbf{\ Primary transitons (tr and lb). }

Primary \textit{transitons }represent intermediate transition states between
(\textit{a}) and (\textit{b}) modes of primary effectons - translational and
librational. Primary transitons (tr and lb) - are radiating or absorbing
photons corresponding to translational and librational bands in oscillatory
spectra. The volumes of primary transitons and primary effectons coincides
(see Table 1).

\textbf{\ Primary electromagnetic and acoustic deformons (tr, lb). }

Electromagnetic primary deformons are a new type of quasiparticles
(excitations) representing a$\,\,3D$ superposition of three standing
electromagnetic waves. The IR\ photons originate and annihilate as a result of
$(a\Leftrightarrow b)_{tr,lb}$ transitions of primary effectons, i.e.
corresponding primary transitons. Such quantum transitions are not accompanied
by the fluctuation of density but with the change of polarizability and dipole
moment of molecules only. Electromagnetic deformons appear as a result of
interception - superposition of 3 pairs of IR photons, penetrating in matter
in different selected directions (1,2,3). We assume, that each of these 3
pairs of photons form a standing wave in the volume of condensed matter.

The linear dimension of each of three edges of primary deformon is determined
by the wave length of 3 intercepting standing photons:%

\begin{equation}
\lambda^{1,2,3}=[(n\tilde{\nu})^{-1}]_{tr,lb}^{1,2,3} \tag{2.5}%
\end{equation}

where:$\,\,\,\,n$ \ is the refraction index and\ $\ (\tilde{\nu})_{tr,lb}$ -
the wave number of translational or librational band. These quasiparticles as
the biggest ones, are responsible for the long-range space-time correlation in
liquids and solids.

In the case when $(b\rightarrow a)_{tr,lb}$ transitions of primary effectons
are accompanied by fluctuation of density they may be followed by emission of
phonons instead of photons. It happened when primary effectons are involved in
the volume of macro- and supereffectons (see below). Primary \textit{acoustic}
deformons may originate or annihilate in such a way. But for primary effectons
the probability of collective spontaneous emission of photons during
$(b\rightarrow a)_{tr,lb}$ transition is much higher than that of phonons.

The coherent electromagnetic radiation as a result of self- correlation of
many dipole moments in composition of coherent cluster, like primary
effectons, containing $\mathbf{N}\gg1$ molecules is already known as
\textbf{superradiance} (Dicke, 1954).

\textbf{\ Secondary effectons (tr and lb)}\textit{. }

In contrast to primary effectons, this type of quasiparticles is
\textit{conventional}. They are the result of averaging of the frequencies and
energies of the ''acoustic'' ($\overline{\mathit{a}}$) and ''optical''
($\overline{\mathit{b}}$) states of effectons with packing numbers $n_{P}>0$,
having the resulting impulse more than zero.

For averaging the energies of such states the Bose-Einstein distribution was
used under the condition when $T<T_{0}\,\,(T_{0}$ is temperature of
degeneration and, simultaneously, temperature of first order phase
transition). Under this condition the chemical potential: $\,\mu=0$ \thinspace
and distribution has a form of Plank equation.

\textbf{\ Secondary transitons (tr and lb). }

Secondary transitons, like primary ones are intermediate transition state
between ($\mathit{\bar{a}}$) and ($\mathit{\bar{b}}$) states of secondary
effecton - translational and librational. Secondary transitons are responsible
for radiation and absorption of phonons. As well as secondary effectons, these
quasiparticles are conventional, i.e. a result of averaging. The effective
volumes of secondary transitons and secondary effectons - coincides.

\textbf{\ Secondary ''acoustic'' deformons (tr and lb). }

This type of quasiparticles is also conventional as a result of 3D
superposition of averaged thermal phonons. These conventional phonons
originate and annihilate in a process of $(\,\bar{a}\,\Leftrightarrow\bar
{b})_{1,2,3}$ thermoactivated transitions of secondary conventional effectons
(translational and librational transitons). In the case of secondary
transitons $(\,\bar{a}\,\Leftrightarrow\bar{b})_{tr,lb}$ transitions are
accompanied by the fluctuation of density.

\textbf{\ Convertons }$\mathbf{(}\mathbf{t}\mathbf{r}\Leftrightarrow
\mathbf{l}\mathbf{b}\mathbf{)}$\textbf{.}

These important excitations are introduced in our model as interconversions
between translational and librational primary effectons. The \textit{(acon)
convertons }correspond to transitions between the (a) states of these
effectons and \textit{(bcon) convertons }- to that between their (b)-states.
As far the dimensions of translational primary effectons are much less than
that of librational ones, the \textit{convertons }could be considered as the
dissociation and association of the primary librational effectons (coherent
clusters). Both of convertons: (\textit{acon}) and (\textit{bcon}) are
accompanied by density fluctuation, inducing phonons with corresponding
frequency in the surrounding medium.

\textbf{\ The ca- and cb- deformons, induced by convertons. }

Three\textit{-}dimensional (3D) superposition of phonons, irradiated by two
types of convertons: \textit{acon }and \textit{bcon}, represent in our model
the acoustic \textit{ca- and cb-deformons. }They have the properties, similar
to that of \textit{secondary deformons, }discussed above.

\textbf{\ The c-Macrotransitons (Macroconvertons) and c-Macrodeformons. }

Simultaneous excitation of the \textit{acon }and \textit{bcon }types of
convertons in the volume of primary librational effectons lead to origination
of big fluctuation, like cavitational one, termed \textit{c-Macrotransitons or
Macroconvertons. }In turn, such fluctuations induce in surrounding medium high
frequency thermal phonons. The 3D-superposition these phonons forms \textit{c-
Macrodeformons. }

\textbf{\ Macroeffectons (tr and lb). }

\textit{Macroeffectons }(A and B) are collective \textit{simultaneous
}excitations of the primary and secondary effectons in the $[A\sim
(a,\,\bar{a})]_{tr,lb}$ \thinspace and\thinspace$\lbrack B\sim(b,\,\bar
{b})]_{tr,lb}$ states in the volume of primary electromagnetic translational
and librational deformons, respectively. This correlation of similar primary
and secondary states results in significant deviations from thermal
equilibrium. The A and B states of macroeffectons (tr and lb) may be
considered as the most probable volume-orchestrated thermal fluctuations of
condensed matter.

\textbf{\ Macrodeformons or macrotransitons (tr and lb). }

This type of conventional quasiparticles is considered in our model as an
\textit{intermediate }transition state of macroeffectons. The $(A\rightarrow
B)_{tr,lb}$ and $(B\rightarrow A)_{tr,lb}$ transitions are represented by the
coherent transitions of primary and secondary effectons \textit{in the volume
of primary electromagnetic deformons} - translational and librational. The
$(A\rightarrow B)_{tr,lb}$ transition of macroeffecton is accompanied by
simultaneous absorption of 3 pairs of photons and that of phonons in the form
of electromagnetic deformons. If $(B\rightarrow A)_{tr,lb}$ transition occur
without emission of photons, then all the energy of the excited B-state is
transmitted into the energy of fluctuation of density and entropy of
Macroeffecton as an isolated mesosystem. It is a dissipative process:
transition from the more ordered structure of matter to the less one, termed
Macrodeformons. The big fluctuations of density during $(A\Leftrightarrow
B)_{tr,lb}$ transitions of macroeffectons, i.e. macrodeformons are responsible
for the Raleigh central component in Brillouin spectra of light scattering
(http://arXiv.org/abs/physics/0003070). Translational and librational
macrodeformons are also related to the corresponding types of viscosity
(http://arXiv.org/abs/physics/0003074). The volumes of
macrotransitons/macrodeformons (tr or lb) and macroeffectons coincide with
that of \textbf{tr} or \textbf{lb primary electromagnetic deformons
correspondingly}.

\textbf{\ Supereffectons. }

This mixed type of conventional quasiparticles is composed of translational
and librational macroeffectons correlated in space and time in the volumes of
superimposed electromagnetic primary deformons (translational and librational
- simultaneously). Like macroeffectons, supereffectons may exist in the ground
$(A_{S}^{*})$ and excited $(B_{S}^{*})$ states representing strong deviations
from thermal equilibrium state.

\textbf{\ Superdeformons or supertransitons. }

This collective excitations have the lowest probability as compared to other
quasiparticles of our model. Like macrodeformons, superdeformons represent the
intermediate $(A^{*}_{S} \Leftrightarrow B^{*}_{S})$ transition state of
supereffectons. In the course of these transitions the translational and
librational macroeffectons undergo simultaneous%

\[
\lbrack(A\Leftrightarrow B)_{tr}\,\,\,\,and\,\,\,(A\Leftrightarrow
B)_{lb}]\text{ \ \ transitions. }
\]
The $(A_{S}^{*}\rightarrow B_{S}^{*})$ transition of supereffecton may be
accompanied by the absorption of two electromagnetic deformons - translational
and librational simultaneously. The reverse $(B_{S}^{*}\rightarrow A_{S}^{*})$
relaxation may occur without photon radiation. In this case the big
\textbf{cavitational fluctuation} originates.

Such a process plays an important role in the processes of sublimation,
evaporation and boiling.

The equilibrium dissociation constant of the reaction:%

\begin{equation}
H_{2}O\rightleftharpoons H^{+}+HO^{-} \tag{2.6}%
\end{equation}
should be related with equilibrium constant of supertransitons: $K_{B_{S}%
^{\ast}\rightleftharpoons A_{S}^{\ast}}$. The $A_{S}^{\ast}\rightarrow
B_{S}^{\ast}$ cavitational fluctuation of supereffectons can be accompanied by
the activation of reversible dissociation of part of water molecules.

\smallskip

\textbf{In contrast to primary and secondary transitons and deformons, the
notions of [macro- and supertransitons] and [macro- and superdeformons]
coincide.} Such types of \textit{transitons and deformons }represent the
dynamic processes in the same volumes of corresponding primary electromagnetic deformons.

Considering the transitions of all types of \textit{translational }deformons
(primary, secondary and macrodeformons), one must keep in mind that the
\textit{librational }type of modes remains the same. And vice versa, in case
of librational deformons, translational modes remain unchanged. Only the
realization of a \textit{convertons and supereffectons are accompanied by the
interconversions between the translational and librational modes, between
translational and librational effectons. }

\smallskip

\textbf{Interrelation between quasiparticles forming solids and liquids. }

Our model includes \textbf{24 }types of quasiparticles (Table. 1):%

\begin{equation}
\left[
\begin{array}
[c]{c}%
4\text{ - \textit{Effectons}}\\
4\text{ - \textit{Transitons}}\\
4\text{ - \textit{Deformons} }%
\end{array}
\right]
\begin{array}
[c]{l}%
\text{translational and librational\textit{,} including }\\
\text{primary and secondary }%
\end{array}
\tag{I}%
\end{equation}%
\begin{equation}
\left[
\begin{array}
[c]{c}%
2\text{ - \textit{Convertons}}\\
2-C\text{\textit{-deformons}}\\
1-Mc\text{\textit{-transiton}}\\
1-Mc\text{\textit{-deformon} }%
\end{array}
\right]
\begin{array}
[c]{l}%
\text{the set of interconvertions }\\
\text{between translational and librational}\\
\text{primary effectons }%
\end{array}
\tag{II}%
\end{equation}%
\begin{equation}
\left[
\begin{array}
[c]{c}%
2\text{ - \textit{Macroeffectons}}\\
2\text{ - \textit{Macrodeformons}}%
\end{array}
\right]
\begin{array}
[c]{l}%
\text{translational and librational }\\
(\text{spatially separated)}%
\end{array}
\tag{III}%
\end{equation}%
\begin{equation}
\left[
\begin{array}
[c]{c}%
1\text{ - \textit{Supereffectons}}\\
1\text{ - \textit{Superdeformons} }%
\end{array}
\right]
\begin{array}
[c]{l}%
\text{translational }\rightleftharpoons\text{ librational}\\
(\text{superposition of \textit{tr} and \textit{lb} effectons}\\
\text{and deformons in the same volume)}%
\end{array}
\tag{IV}%
\end{equation}

\smallskip

\textbf{Each level in the hierarchy }of quasiparticles (I - IV) introduced in
our model is based on the principle of correlation of corresponding type of a
dynamic process in space and time. All of these quasiparticles are constructed
on the same physical principles as 3D -superposition (interception) of
different types of standing waves.

\textbf{Such a system can be handled as an ideal gas of quasiparticles of 24
types. }

As far each of the effecton's types: $tr\,\,\,$and$\,\,\,lb$, macroeffectons
$[tr+lb]$ and supereffectons $[tr/lb]$ has two states (acoustic and optic)
\textbf{the total number of excitations is equal to:}
\[
\mathbf{N}_{ex}=\mathbf{3}\mathbf{1}%
\]

\textbf{Three types of standing waves are included in our model:}

- de Broglie waves of particles (waves B);

- acoustic waves (phonons);

- electromagnetic waves (IR photons).

This classification reflects the \textit{duality of matter and field }and
represent their self-organization and interplay on mesoscopic and macroscopic levels.

Our hierarchical system includes a gradual transition from the \textit{Order
}(primary effectons, transitons and deformons) to the \textit{Chaos }(macro-
and superdeformons). It is important, however, that in accordance with the
model proposed, this thermal \textit{Chaos is ''organized'' }by hierarchical
superposition of definite types of the ordered quantum excitations. It means
that the final dynamics condensed matter only ''looks'' as chaotic one. Our
approach makes it possible to take into account the \textbf{Hidden Order of
Condensed Matter. }

\smallskip

The increasing or decreasing in the concentration of primary deformons is
directly related to the shift of $(a\Leftrightarrow b)_{tr,lb}$ equilibrium of
the primary effectons leftward or rightward, respectively. This shift, in
turn, leads also to corresponding changes in the energies and concentrations
of secondary effectons, deformons and, consequently, to that of super- and
macro-deformons. \textbf{It means the existing of feedback reaction between
subsystems of the effectons and deformons, necessary for long-range
self-organization in macroscopic volumes of condensed matter.}

\begin{center}%
\begin{center}
\includegraphics[
height=4.51in,
width=4.3889in
]%
{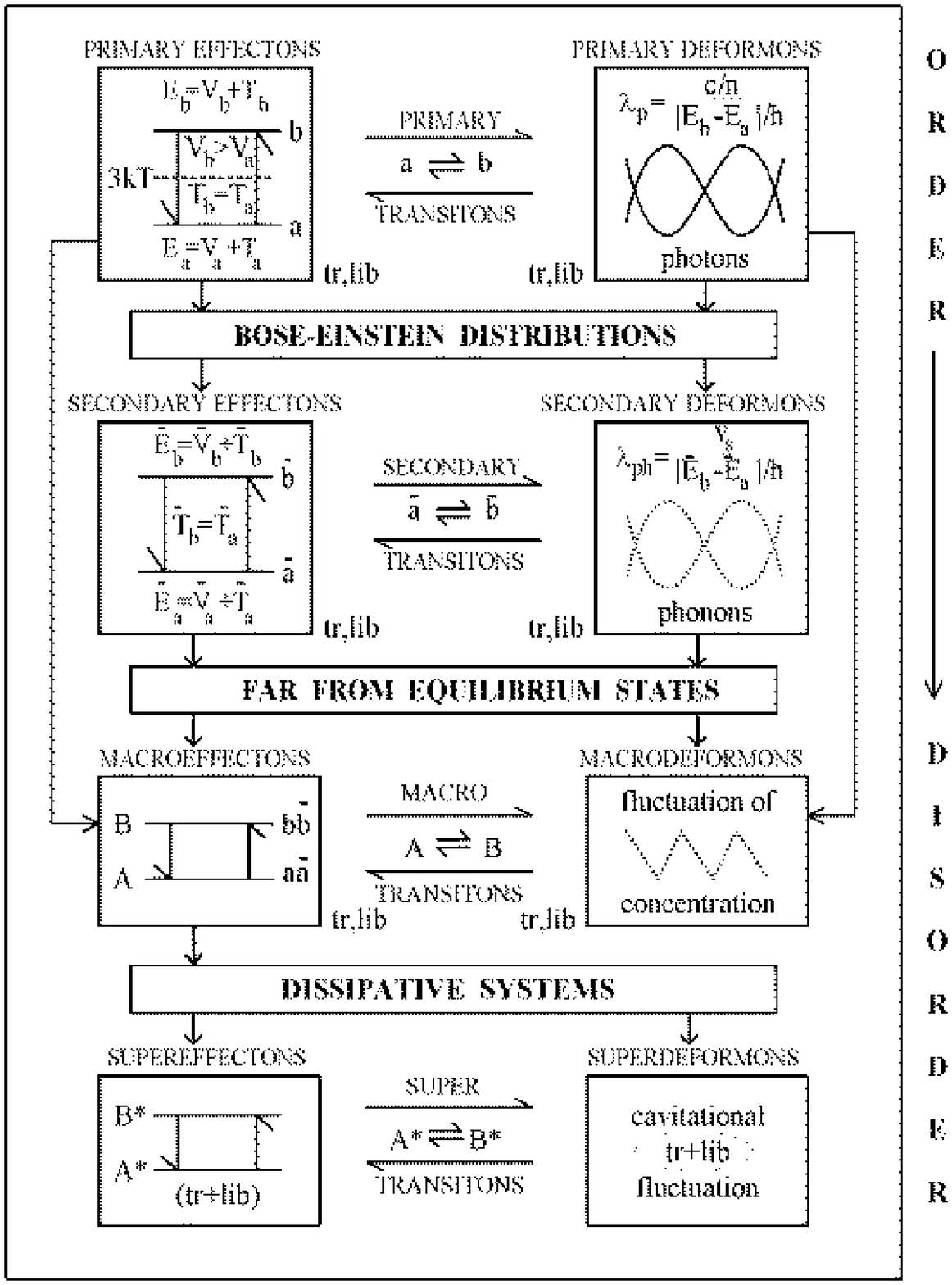}%
\end{center}
\end{center}

\begin{quote}
\textbf{Table~1. }Schematic representation of the 18 types of quasiparticles
of condensed matter as a hierarchical dynamic system, based on the effectons,
transitons and deformons. Total number of \textit{quasiparticles}, introduced
in Hierarchic concept is 24. Six collective excitations, related to
\textit{convertons}- interconversions between primary librational and
translational effectons and their derivatives are not represented here for the
end of simplicity.

\smallskip
\end{quote}

The total internal energy of substance is determined by the contributions of
all types of quasiparticles with due regard for their own energy,
concentration and probability of excitation. Contributions of \textit{super}-
and \textit{macro}effectons and corresponding \textit{super}- and
\textit{macro}deformons as well as \textit{polyeffectons and coherent
superclusters} to the internal energy of matter normally are small, due to
their low probability of excitation, big volume and, consequently, low concentration.

\medskip

The sizes of \textbf{primary} effectons determine the mesoscopic scale of the
condensed matter organization. \textbf{According to our model, the domains,
nods, crystallites, and clusters observed in solid bodies and in liquid
crystals, polymers and biopolymers - can be a consequence of primary
translational or librational effectons. \smallskip}

\begin{center}
\textbf{III. }\ \textbf{HIERARCHIC THERMODYNAMICS}{\Large \ }
\end{center}

\smallskip

\begin{center}
\textbf{3.1. The internal energy of matter as a hierarchical system of
collective excitations }
\end{center}

\smallskip

The quantum theory of crystal heat capacity leads to the following equation
for the density of thermal internal energy (Ashkroft, Mermin 1976):%

\begin{equation}
\epsilon={\frac{1}{V}}{\frac{{\sum\,}E_{i}\exp(-E_{i}/kT)}{{\sum\,}\exp
(-E_{i}/kT)}} \tag{3.1}%
\end{equation}
where V - the crystal volume;\thinspace$E_{i}$ - the energy of the
i-stationary state.

According to our Hierarchic theory, the internal energy of matter is
determined by the concentration $(n_{i})$ of each type of quasiparticles,
probabilities of excitation of each of their states $(P_{i})$ and the energies
of corresponding states $(E_{i})$. The condensed matter is considered as an
''ideal gas'' of 3D standing waves of different types (quasiparticles and
collective excitations). However, the dynamic equilibrium between types of
quasiparticles is very sensitive to the external and internal perturbations.

The total partition function - the sum of the relative probabilities of
excitation of all states of quasiparticles (the resulting
\textit{thermoaccessibility}) is equal to:%

\begin{align}
Z  &  =\sum_{tr,lb}\left\{
\begin{array}
[c]{c}%
\left(  P_{ef}^{a}+P_{ef}^{b}+P_{d}\right)  +\\
+\left(  \bar{P}_{ef}^{a}+\bar{P}_{ef}^{b}+\bar{P}_{d}\right)  +\\
+\left[  \left(  P_{M}^{A}+P_{M}^{B}\right)  +P_{D}^{M}\right]
\end{array}
\right\}  _{tr,lb}+\tag{3.2}\\
&  +\left(  P_{ac}+P_{bc}+P_{\text{cMd}}\right)  +\left(  P_{S}^{A}+P_{S}%
^{B}+P_{D^{\ast}}^{s}\right) \nonumber
\end{align}
Here we take into account that the probabilities of excitation of primary and
secondary transitons and deformons are the same $(P_{d}=P_{t};\;\;\bar{P}%
_{d}=\bar{P}_{t})$ and related to the same processes:%

\[
(a\Leftrightarrow b)_{tr,lb}\text{ \ \ and \ \ }(\bar{a}\Leftrightarrow\bar
{b})_{tr,lb}\,\,\text{ transitions. }%
\]
The analogous situation is with probabilities of \textit{a, b and cM
convertons }and corresponding acoustic deformons excitations: $P_{ac}%
,\,P_{bc}\;$ and \ $P_{\text{cMd}}=P_{\text{cMt}}$. So it is a reason for
taking them into account in the partition function only ones.

The final formula for the total internal energy of $(U_{\text{tot}})$ of one
mole of matter leading from mesoscopic model, considering the system of 3D
standing waves as an ideal gas is:%

\[
U_{\text{tot}}=V_{0}{\frac{1}{Z}}\sum_{tr,lb}\left\{  \left[
\begin{array}
[c]{c}%
n_{ef}\left(
\begin{array}
[c]{c}%
P_{ef}^{a}E_{ef}^{a}+P_{ef}^{b}E_{ef}^{b}+P_{t}E_{t}%
\end{array}
\right) \\
+n_{d}P_{d}E_{d}%
\end{array}
\right]  \right.  +
\]%
\[
+\left[  \bar{n}_{ef}\left(
\begin{array}
[c]{c}%
\bar{P}_{ef}^{a}\bar{E}_{ef}^{a}+\bar{P}_{ef}^{b}\bar{E}_{ef}^{b}+\bar{P}%
_{t}\bar{E}_{t}%
\end{array}
\right)  +\bar{n}_{d}\bar{P}_{d}\bar{E}_{d}\right]  +
\]%
\[
+\left[  n_{M}\left(
\begin{array}
[c]{c}%
P_{M}^{A}E_{M}^{A}+P_{M}^{B}E_{M}^{B}%
\end{array}
\right)  +n_{D}P_{M}^{D}E_{M}^{D}\right]  _{tr,lb}+
\]%
\[
+V_{0}{\frac{1}{Z}}\left[  n_{\text{con}}\left(
\begin{array}
[c]{c}%
P_{ac}E_{ac}+P_{bc}E_{bc}+P_{\text{cMt}}E_{\text{cMt}}%
\end{array}
\right)  \right.  +
\]%
\[
+\left.
\begin{array}
[c]{r}%
\left(  n_{\text{cda}}P_{ac}E_{ac}+n_{\text{cdb}}P_{bc}E_{bc}+n_{\text{cMd}%
}P_{\text{cMd}}E_{\text{cMd}}\right)
\end{array}
\right]  +
\]%
\begin{equation}
+V_{0}{\frac{1}{Z}}n_{s}\left[
\begin{array}
[c]{c}%
\left(
\begin{array}
[c]{c}%
P_{S}^{A^{\ast}}E_{S}^{A^{\ast}}+P_{S}^{B^{\ast}}E_{S}^{B^{\ast}}%
\end{array}
\right)  {\ }+n_{D^{\ast}}P_{S}^{D^{\ast}}E_{S}^{D^{\ast}}%
\end{array}
\right]  \tag{3.3}%
\end{equation}
The contributions of the all types the effecton's, transitons, convertons and
deformons in total internal energy may be calculated separately as a parts of
(3.3). It is shown, that all similar contributions to the total kinetic energy
can be also simulated. The corresponding potential energy of quantum
excitations is a result of difference between total and kinetic energies.

The small contribution of intramolecular dynamics ($U_{\text{in}})$ to
$U_{\text{tot}}$, related to energy of fundamental molecular modes $(\nu
_{p}^{i})$, may be estimated using Plank distribution.

It has been shown by our computer simulations for the case of water and ice
that $U_{\text{in}}\ll U_{\text{tot}}$. It should be general condition for any
condensed matter.\smallskip

\textbf{For} \textbf{the meaning of the variables in formulae (3.2
-\thinspace3.4), necessary for the internal energy calculations, see full
version of paper, presented at the Archives of Los -Alamos: http://arXiv.org/abs/physics/0003044}

$\mathbf{\smallskip}$

\begin{center}
\textbf{IV.\ \ QUANTITATIVE VERIFICATION OF HIERARCHIC THEORY }

\textbf{ON\ EXAMPLES OF ICE AND WATER\medskip}
\end{center}

All the calculations, based on Hierarchic mesoscopic concept, were performed
on the personal computers. \textbf{The special software: ''Comprehensive
analyzer of matter properties'' was elaborated (copyright, 1997, Kaivarainen).
The program allows to evaluate more than three hundred parameters of any
condensed matter if the following data are available in the temperature
interval of interest: }

\textbf{1.}~\textbf{Positions of translational and librational bands in
oscillatory spectra; }

\textbf{2. Sound velocity; }

\textbf{3. Molar volume or density; }

\textbf{4. Refraction index}.\smallskip

The temperature dependences of different parameters for ice and water,
computed using the formulas of our mesoscopic theory, are presented in
Figs.(1-4). It is only a small part of available information. In principle, it
is possible to calculate about 300 different parameters for liquid and solid
state of any condensed matter (Kaivarainen, 1995 and
http://arXiv.org/abs/physics/0003044). \medskip

\begin{center}
\textbf{4.1. Discussion of theoretical temperature dependences and comparison
with experiment }\medskip
\end{center}

It will be shown below that our hierarchic theory makes it possible to
calculate a lot of parameters for water and ice. Those of them that could be
measured experimentally are in excellent correspondence with results of theory.

On lowering down the temperature the total internal energy of ice (Fig. 1a)
and its components decreases nonlinearly with temperature coming closer to
absolute zero. The same parameters for water are decreasing almost linearly
within the interval $(100-0)^{0}C\;($Fig. 1b).

In computer calculations, the values of $C_{p}(t)$ can be determined by
differentiating $U_{\text{tot}}$ numerically at any of temperature interval.

\begin{center}%
\begin{center}
\includegraphics[
height=2.4189in,
width=4.6985in
]%
{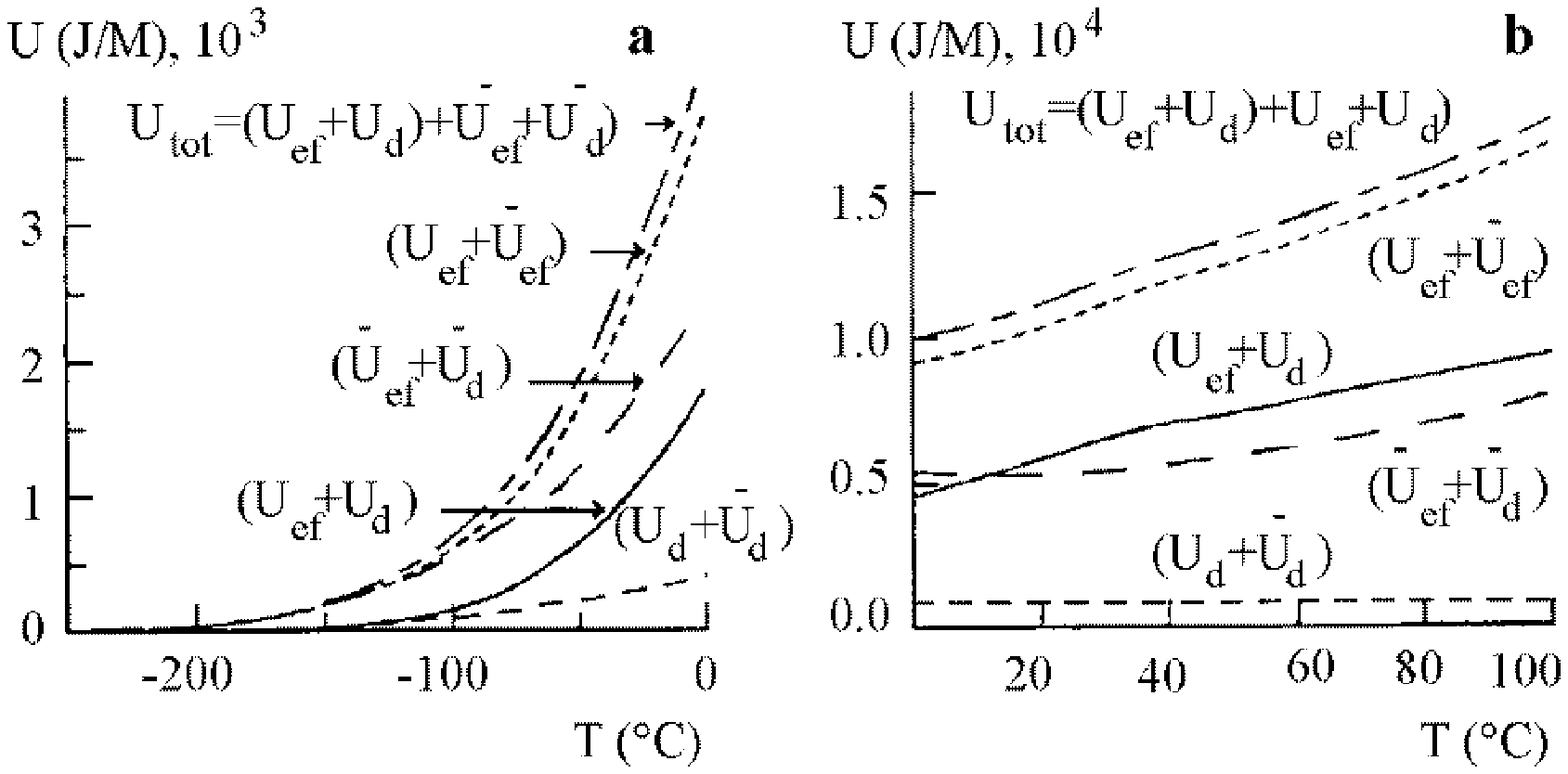}%
\end{center}
\end{center}

\begin{quotation}
\textbf{Fig.~1}. (a,b).~Temperature dependences of the total internal energy
$(U_{\text{tot}})$ and different contributions for ice (a) and water (b) (eqs.
4.3 - 4.5). Following contributions to $U_{\text{tot}}$ are presented:

$(U_{ef}+\bar{U}_{ef})$ is the contribution of primary and secondary effectons;

$(U_{d}+\bar{U}_{d})$ is the contribution of primary and secondary deformons;

$(U_{ef}+U_{d})$ is the contribution of primary effectons and deformons;

$(\bar{U}_{ef}+\bar{U}_{d})$ is the contribution of secondary effectons and deformons.

The contributions of macro- and supereffectons to the total internal energy
and corresponding macro- and superdeformons, as well as all types of
convertons, are much smaller than those of primary and secondary effectons and
deformons.\medskip\ 
\end{quotation}

It follows from Fig. 2a that the mean value of heat capacity for ice in the
interval from -75 to $0^{o}C$ is equal to:%

\begin{equation}
\bar{C}_{p}^{ice}={\frac{\Delta U_{\text{tot}}}{\Delta T}}\approx39\,J/M\cdot
K=9.3\text{ cal}/M\cdot K \tag{4.1}%
\end{equation}
For water within the whole range $\Delta T=100^{0}C$, the change in the
internal energy is: $\Delta U=17-9.7=7.3\,kJ/M\;($Fig.2b). This corresponds to
mean value of heat capacity of water:
\begin{equation}
C_{p}^{water}={\frac{\Delta U_{\text{tot}}}{\Delta T}=73\,J/M\cdot
K=17.5\,cal/M\cdot K} \tag{4.2}%
\end{equation}

These results of our theory agree well with the experimental mean values
$C_{p}=18$ Cal$/M\cdot K$ for water and $\;C_{p}=9\,cal/M\cdot K$ for ice.\smallskip

\begin{center}
\textbf{ }\smallskip

\textbf{4.2. Temperature dependence of properties of primary librational
effectons}\smallskip
\end{center}

The number of $H_{2}O$ molecules within the \textbf{primary libration
effectons }of water, which could be approximated by a cube, decreases
from\thinspace$\;n_{M}=$ 280 at $0^{0}$ \thinspace to\ \thinspace$n_{M}$
$\simeq3$ at $100^{0}\;($Fig. 2a). It should be noted that at physiological
temperatures $(35-40^{0})$ such quasiparticles contain nearly 40 water
molecules. This number is close to that of water molecules that can be
enclosed in the open interdomain protein cavities judging from X-ray data. The
\textit{flickering} of these clusters, i.e. their $\left(
dissociation\,\rightleftharpoons\,association\right)  $ due to
$[lb\Leftrightarrow tr]$ conversions in accordance with our model is directly
related to the large-scale dynamics of proteins.

\begin{center}%
\begin{center}
\includegraphics[
height=2.3186in,
width=4.6138in
]%
{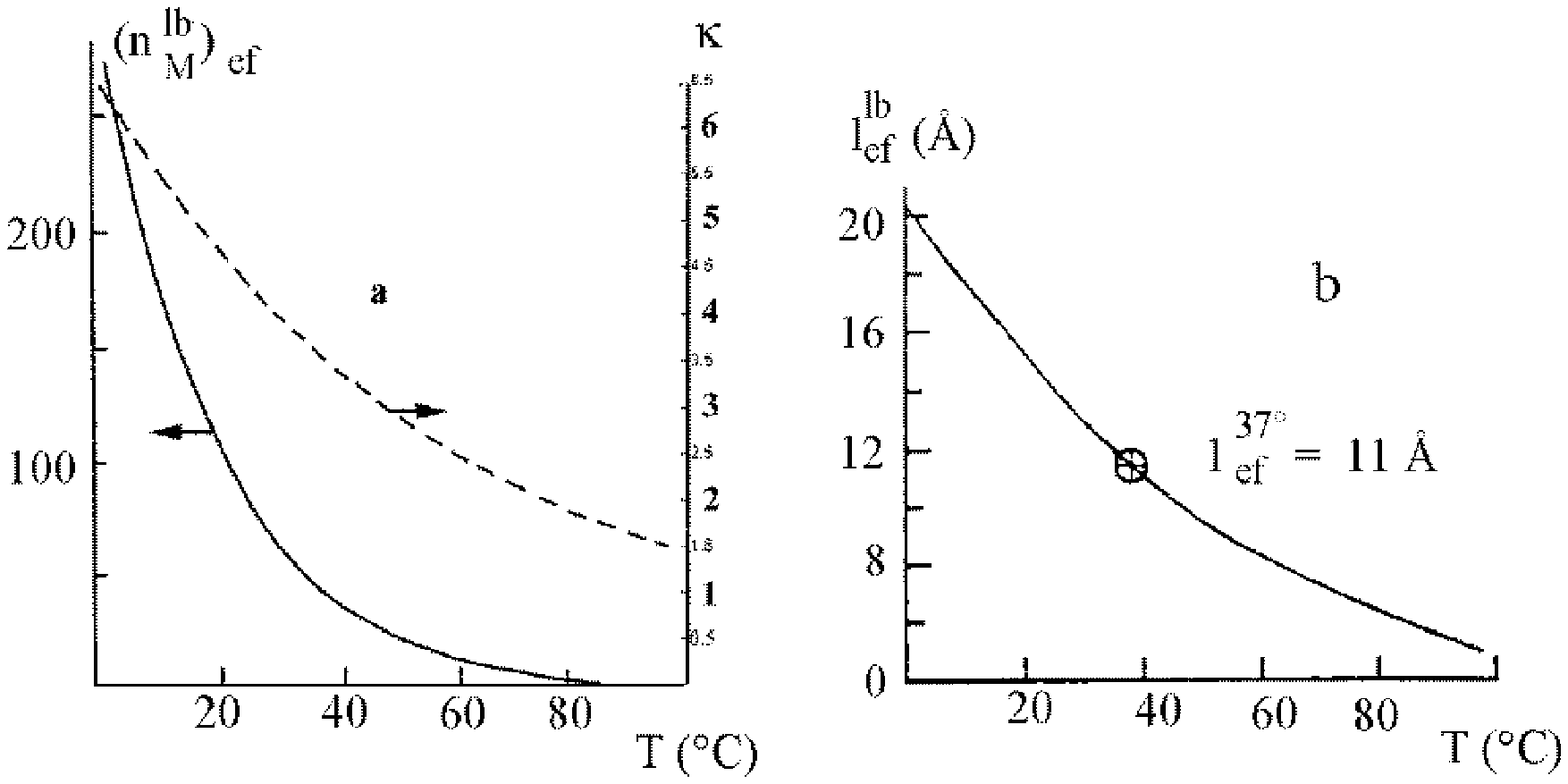}%
\end{center}
\end{center}

\begin{quotation}
\textbf{Fig.}$\mathbf{\ }$2$\mathbf{.\;}(a):\;$The temperature dependencies of
the number of $H_{2}O$ molecules in the volume of primary librational effecton
$(n_{M}^{lb})_{ef},\;$left axis) and the number of $H_{2}O$ per length of this
effecton edge $(\kappa$, right axis); \thinspace(b): the temperature
dependence of the water primary librational effecton (approximated by cube)
edge length $[l_{ef}^{\text{lib}}=\kappa(V_{0}/N_{0})^{1/3}]$.

\smallskip
\end{quotation}

\textit{It is very important that the linear dimensions of such water clusters
(11\AA) at physiological temperature are close to common ones for protein
domains }(Fig. 2b).

\textbf{Such spatial correlations indicate that the properties of water
exerted a strong influence on the biological evolution of macromolecules,
namely, their dimensions and allosteric properties due to cooperativeness of
intersubunit water clusters.}

\smallskip

\ We assume here that integer and half-integer values of number of water
molecules per effecton's edge $\left[  \kappa\right]  $ (Fig. 2a) reflect the
conditions of increased and decreased stabilities of water structure
correspondingly. It is apparently related to the stability of primary
librational effectons as cooperative and coherent water clusters.

Nonmonotonic behavior of water properties with temperature is widely known and
well confirmed experimental fact (Drost-Hansen, 1976, 1992; Clegg and
Drost-Hansen, 1991; Etzler, 1991; Roberts and Wang, 1993; Roberts and Wang,
1993; Roberts, et al., 1993, 1994; Wang et al., 1994).

We can explain this interesting and important for biological functions
phenomenon because of \textbf{competition between two factors: quantum and
structural ones in stability of primary librational effectons.} \textit{The
quantum factor such as wave B length}, determining the value of the effecton
edge:
\begin{equation}
\left[  l_{ef}=\kappa(V_{0}/N_{0})^{1/3}\;\symbol{126}\;\lambda_{B}\right]
_{lb} \tag{4.3}%
\end{equation}
decreases monotonously with temperature increasing. The \textit{structural
factor} is a discrete parameter depending on the water molecules effective
length: $\;l_{H_{2}O}=(V_{0}/N_{0})^{1/3}\;$ and their number $\left[
\kappa\right]  $ in the effecton's edge.

\textbf{In accordance with our model, the shape of primary librational
effectons in liquids and of primary translational effectons in solids could be
approximated by parallelepiped in general case or by cube, when corresponding
thermal movements of molecules (lb and/or \thinspace tr) and are isotropic.}

We suggest that when $(l_{ef})$ corresponds to \textit{integer} number of
$H_{2}O$, i.e.
\begin{equation}
\left[  \kappa=(l_{ef}/l_{H_{2}O})=2,3,4,5,6...\right]  _{lb} \tag{4.4}%
\end{equation}
the \textit{competition} between quantum and structural factors is minimum and
primary librational effectons are most stable. On the other hand, when
$(l_{ef}/l_{H_{2}O})_{lb}$ is half-integer, the librational effectons are less
stable (the \textit{competition is maximum}). In the latter case
$(a\Leftrightarrow b)_{lb}$ equilibrium of the effectons must be shifted
rightward - to less stable state of these coherent water clusters.
Consequently, the probability of dissociation of librational effectons to a
number of much smaller translational effecton, i.e. probability of [lb/tr]
convertons increases and concentration of primary librational effectons
decreases. Experimentally the nonmonotonic change of this probability with
temperature could be registered by dielectric permittivity, refraction index
measurements and by that of average density of water. The refraction index
change should lead to corresponding variations of surface tension, vapor
pressure, viscosity, self-diffusion in accordance to our hierarchic theory
(Kaivarainen, 1995). \smallskip

\begin{center}
\textbf{4.3. Mechanism of first order phase transition in terms of hierarchic
theory}\textit{\ }\smallskip
\end{center}

The mechanism of 1st order phase transition - water freezing, leading from our
Hierarchic theory, is important for understanding the mechanism of antifreeze
molecules action.

\ The abrupt increase of the total internal energy (U) as a result of ice
melting (Fig. 3a), equal to 6.27 kJ/M, calculated from our theory is close to
the experimental data (6 kJ/M)$\;$(Eisenberg,\ 1969). The resulting
thermoaccessibility or partition function (Z, eq.3.2) during $\left[
ice\rightarrow water\right]  $ transition decreases abruptly, while potential
and kinetic energies increase (Fig. 3b).%

\begin{center}
\includegraphics[
height=1.8775in,
width=4.0957in
]%
{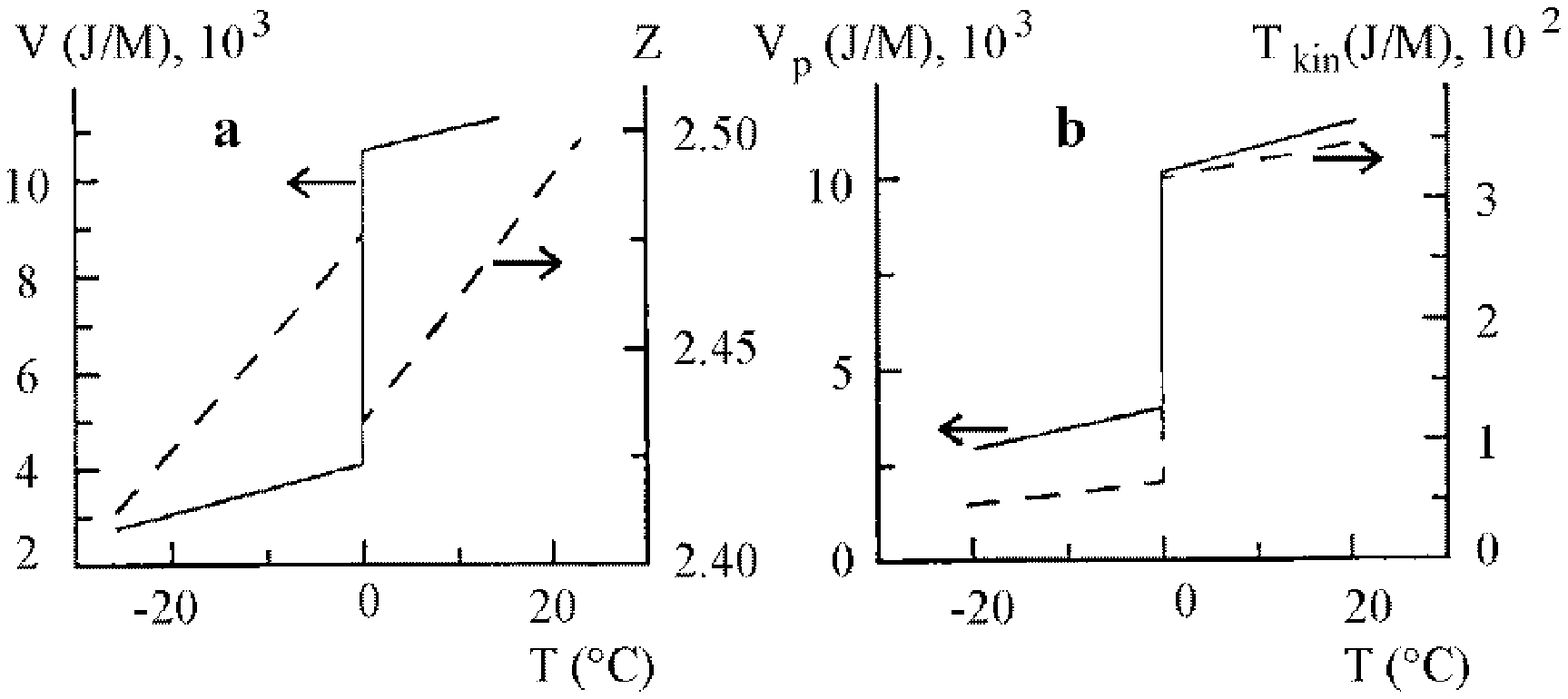}%
\end{center}

\begin{quotation}
\textbf{Fig.~3. } The total internal energy $(U=T_{\text{kin}}+V_{p})$ change
during ice-water phase transition and change of the resulting
thermoaccessibility (Z) - (a); changes in kinetic $(T_{\text{kin}})$ and
potential $(V_{p})$ energies (b) as a result of the same transition.

\smallskip
\end{quotation}

It is important that at the melting point the $H_{2}O$ molecules number in a
\textit{primary translational effecton }$(n_{M}^{tr})_{ef}$ decreases from 1
to $\simeq0.4\;($Fig. 4a). It means that \textit{the volume of this
quasiparticle type gets smaller than the volume occupied by} $H_{2}O$
\textit{molecule.} According to our model, under such conditions the
individual water molecules get the independent translation mobility,
increasing the number of \emph{tr }degrees of freedom. The number of water
molecules forming a \textit{primary libration effecton} decreases abruptly
from about 3000 to 280, as a result of melting. This also increases the number
of molecules, participating in translations. The number of $H_{2}O$ in the
secondary librational effecton decreases correspondingly from $\sim1.25$ to
0.5 (Fig. 4b).

Fig. 5 a,b contains more detailed information on changes in primary
librational effecton parameters in the course of ice melting.

The theoretical dependences obtained allow us to give a clear interpretation
of the first order phase transitions. The condition of melting at $T=T_{cr}$
is realized in the course of heating when the number of molecules in the
volume of primary translational effectons $n_{M}$ decreases:%

\begin{equation}
n_{M}\geq1(T\leq T_{cr})\overset{T_{c}}{\rightarrow}\text{ }n_{M}\leq1(T\geq
T_{cr}) \tag{4.5}%
\end{equation}

\begin{center}%
\begin{center}
\includegraphics[
height=2.0704in,
width=4.2903in
]%
{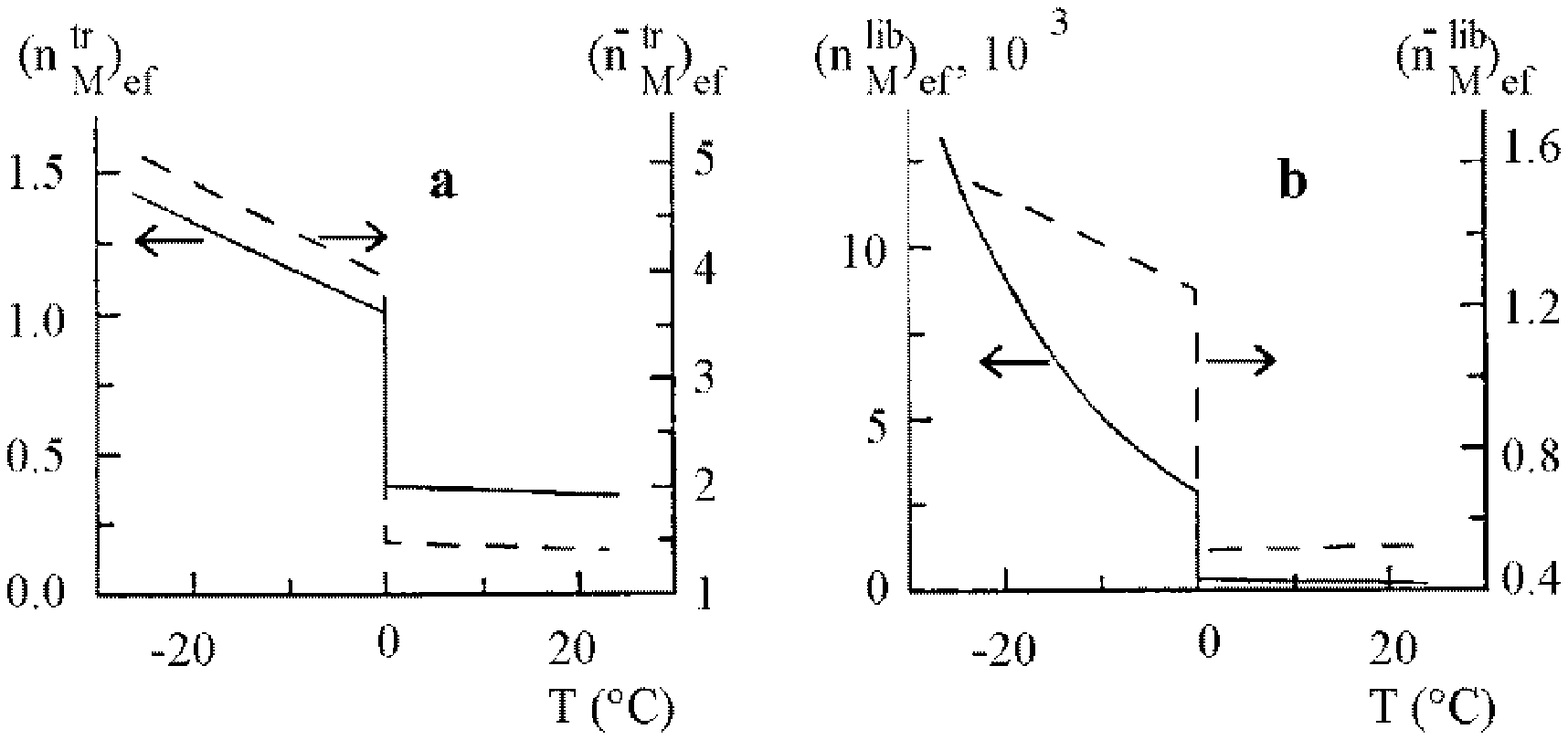}%
\end{center}
\end{center}

\begin{quotation}
\textbf{Fig.~4. }Changes of the number of $H_{2}O$ molecules forming primary
$(n_{M}^{tr})_{ef}$ and secondary $(\bar{n}_{M}^{tr})_{ef}$ translational
effectons during ice-water phase transition (a). Changes in the number of
$H_{2}O$ molecules forming primary $(n_{M}^{\text{lb}})_{ef}$ and secondary
$(\bar{n}_{M}^{\text{lb}})_{ef}$ librational effectons (b) as a result of
phase transitions.

\smallskip
\end{quotation}

The process of boiling, i.e. [liquid $\rightarrow$ gas] transition, as seen
from Fig. 4a, is also determined by condition (4.5), but in this case it is
realized for primary \textit{librational }effectons.

In other words this means that [gas $\rightarrow$ liquid] transition is
related to origination (condensation) of the \textit{primary librational
effectons }which contain more than one molecule of substance.

\textbf{In a liquid as compared to gas, the quantity of rotational degrees of
freedom is decreased due to librational coherent effectons formation, but the
number of translational degrees of freedom remains the same. The latter, in
turn also decreases during [liquid }$\rightarrow$\textbf{\ solid] phase
transition, when the wave B length of molecules corresponding to their
translations begins to exceed the mean distances between the centers of
molecules (Fig. 4a). }This process is accompanied by partial Bose-condensation
(BC) of translational waves B and by the formation of coherent primary
translational effectons, including more than one molecule. The size of
librational effectons grows up strongly during this $\left[  water\rightarrow
ice\right]  $ transition. This enlarged \emph{lb-}effectons and their
associates may fulfill the role of nucleation centers, necessary for [liquid
$\rightarrow\,$solid] transition. Interaction and adsorption of individual
water molecules with such big structures damp the translational mobility of
water molecules and increase their corresponding kind of de Broglie wave
length till critical value, allowing mesoscopic molecular BC.

Consequently, it follows from the theory, that all factors, preventing the
enlargement of primary [lb] effectons in the range of molecular numbers:
280-3000 and their possible polymerization will serve as an antifreeze factors.

\begin{center}%
\begin{center}
\includegraphics[
height=2.0418in,
width=4.3024in
]%
{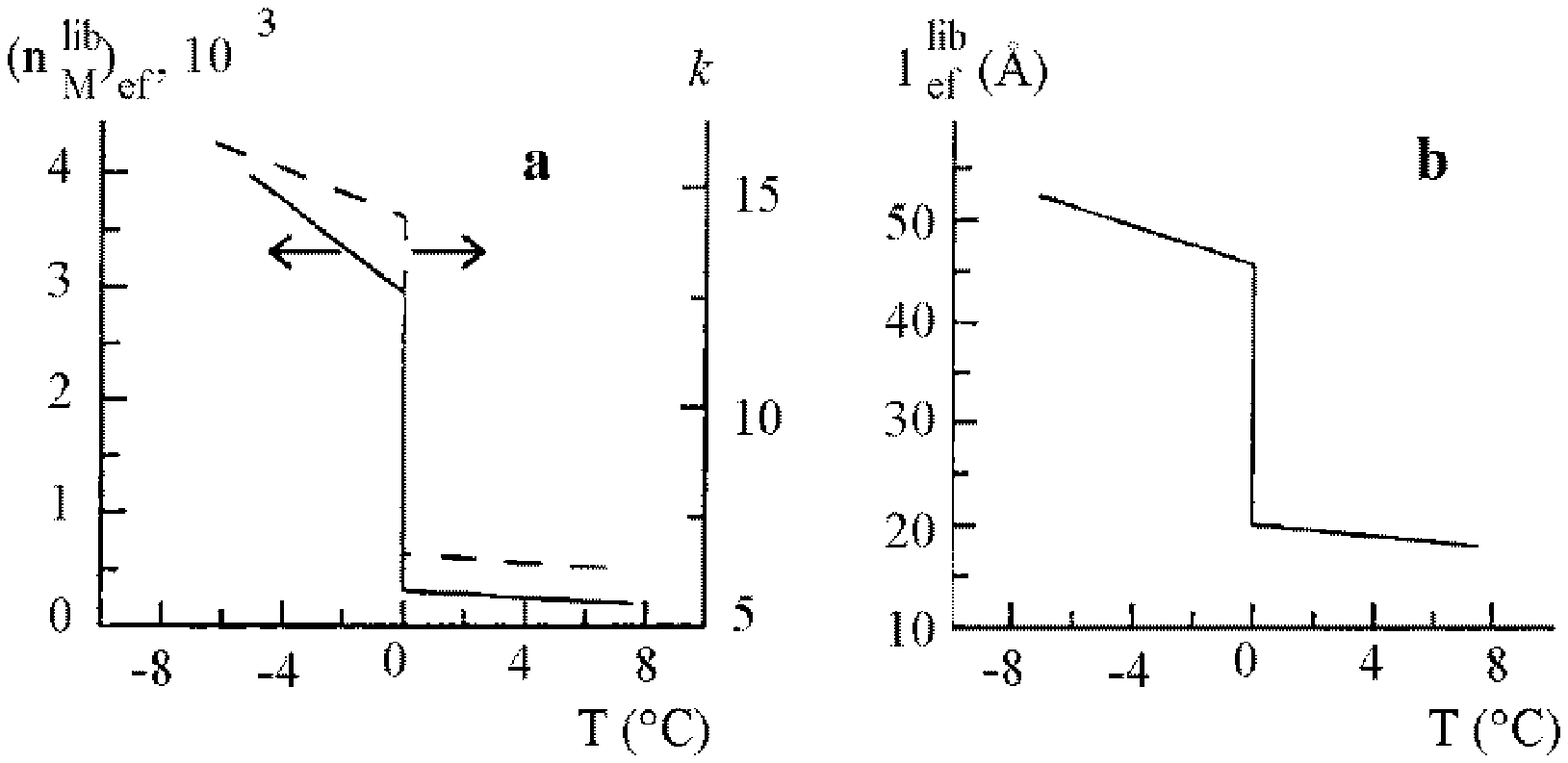}%
\end{center}
\end{center}

\begin{quotation}
\textbf{Fig.~5. }Changes of the number of $H_{2}O$ molecules forming a primary
librational effecton $(n_{M}^{lb})_{ef}$, the number of $H_{2}O$ molecules
$(\kappa)$ in the edge of this effecton (a) and the length of the effecton
edge: $l_{ef}^{lb}=\kappa(V_{0}/N_{0})^{1/3}\;$(b) during the ice-water phase transition.

\smallskip
\end{quotation}

Our results do not contradict to conventional interpretation of first order
phase transitions and give them clear quantum explanation. The number of
molecules in volumes of primary effectons: translational and librational can
be considered as the ''parameters of order'', characterizing the condition of
first order phase transition.

\textbf{The results of\ computer simulations, confirms the correctness of new
theory of condensed matter, as a hierarchic system of 3D standing waves of
different nature and demonstrate its possibilities. \ It may be applied for
understanding the mechanism of antifreeze proteins action.}

\begin{center}

\textbf{V. }\ \textbf{PROPERTIES OF ANTIFREEZE PROTEINS}\medskip

\textbf{5.1. Properties of the antifreeze (AFP) and ice-nucleation proteins
(INP)\medskip}
\end{center}

The antifreeze proteins (AFP), discovered in arctic fish by P. Scholander 1957
and studied later profoundly by group, led by Garth Fletcher and Choy Hew are
widely spread in the nature. AFP are used for surviving at low temperatures
not only by Arctic and Antarctic fish, but also by frogs, insects, plants,
crops, microbes, etc.

The interest to AFP is substantial first of all because of undergoing large
research programs on the development of genetically manipulated crop plants
adapted to cold climate. Quite recently, the importance of AFP has been even
intensified by a range of more straightforward technical applications. For
example, AFP can be used for:

- preservation of blood substances;

- protection of human organs for transplantation at low temperatures;

- preservation of texture of frozen food after melting.

All these applications exploit AFP%
\'{}%
ability to inhibit formation of big ice crystals, damaging the cytoplasmic
supramolecular structure of cells, cell membranes and tissues. However, the
mechanism of specific influence of AFP on primary hydrated shell, vicinal
water and bulk water is still obscure. On the other hand, this understanding
is very important for the full exploitation of AFP .

At the 216th American Chemical Society Meeting in Boston (autumn 1998)
possible mechanisms of action of AFP, including the role of hydrogen bonding
in the interaction of the protein structure with ice, and contribution of
hydrophobic interaction between the protein and water was discussed. The
summary by Dr. Fletcher, president of A/F Protein Inc., was not optimistic:
''We are less certain of how it works now than we were 10 years ago''. Peter
Davies supposed, that specific hydrogen bonds on the surface of protein
inhibit crystal growth. The interesting observation is that AFP accumulates at
the interface between ice and water. A. Haymet (University of Houston)
hypothesized that a hydrophobic reaction between the protein and the
neighboring water prevents water from forming ice crystals. To test this idea
Harding (University of Sydney) synthesized several mutant flounder proteins
and compared them with wild type proteins. Two mutants lacking all threonine
residues - the hydroxyl groups necessary for hydrogen bonding - behaved
exactly like the wild type proteins without hydroxyls. These results mean that
hydrogen [protein-water] bonding does not play the main role in antifreeze
activity (Haymet et al., 1998).

In a series of experiments, Davies and colleagues compared antifreeze proteins
found in fish with much stronger ones produced in some moths and beetles. The
insect proteins inhibited ice growth up to -6 degrees Celsius-about four times
the efficacy of known fish proteins. The results point out, in turn, that
threonine with its hydroxyl groups plays an important role in AFP activity,
since this amino acid is more abundant in insect proteins than in fish proteins.

AFP found in the winter flounder, is structurally long and thin, while the
other three types of AFP are more circular. Therefore, it may be misleading to
use that protein as a model for how all the other types work.

The fact that all antifreeze proteins studied are rich in hydrogen-bonding
amino acid groups cannot be ignored, especially since many other structural
and sequential aspects of antifreeze proteins differ drastically from type to
type. The understanding of general principles of how the different types of
antifreeze mechanisms work could help to build up effective synthetic
molecules for different practical purposes.

AFPs lower the non-equilibrium freezing point of water (in the presence of
ice) below the melting point, thereby producing a difference between the
freezing and melting points that has been termed thermal hysteresis. In
general, the magnitude of the thermal hysteresis depends upon the specific
activity and concentration of AFP. The study of (Li et al., 1998) describes
several low-molecular-mass solutes that enhance the thermal hysteresis
activity of an AFP from overwintering larvae of the beetle \emph{Dendroides
canadensis}. The most active of these is citrate, which increases the thermal
hysteresis nearly sixfold from 1.2
${{}^\circ}$%
C in its absence to 6.8
${{}^\circ}$%
C. Solutes which increase activity approximately fourfold are succinate,
malate, aspartate, glutamate and ammonium sulfate. Glycerol, sorbitol, alanine
and ammonium bicarbonate increase thermal hysteresis approximately threefold.
Sodium sulfate at high concentration (0.5 mol/l) eliminated activity of all
listed solutes. Solute concentrations between 0.25 and 1 mol/1 were generally
required to elicit optimal thermal hysteresis activity. Glycerol is the only
one of these enhancing solutes that is known to be present at these
concentrations in overwintering \emph{D. Canadensis}. The mechanism of this
enhancement of AFP activity by low molecular solutes is unknown. The AFP used
in this study was (DAFP-4) type. The mature protein consists of 71 amino acid
residues arranged in six 12- or 13-mer repeats with a consensus sequence
consisting of Cys-Thr-X3-Ser-X5-X6-Cys-X8-X9-Ala-X11-Thr-X13, where X3 and X11
tend to be charged residues.

For understanding the mechanisms of AFP, it is important to compare them with
ice nucleation proteins (INP) with the opposite effect on water structure.

INP are responsible for another strategy to survive freezing temperature - to
make the animals and plants freeze tolerant. Such organisms secrete INP to the
extracellular environment. INP effects that ice-formation outside the cells
decreases since it minimizes the extent of supercooling, so that freezing
becomes less destructive to cells, making a time for cells to the cold
acclimation. Such adaptation should be accompanied by dehydration of cells due
to decreasing of water activity in the space out of cells and corresponding
passive osmotic diffusion of water from cells.

The tertiary structure of the ice-nucleation protein is likely to be regular,
consistent with the expectation of its forming a template for ice-matching.
INP stimulates the formation of ''ice embryo'' - a region with some
characteristics of macroscopic ice.

INP catalyze ice formation at temperatures much higher than most organic or
inorganic substances (Gurian-Sherman and Lindow, 1993). Models of INPs predict
that they form a planar array of hydrogen bonding groups that are closely
complementary to the ice crystal. The circular dichroism spectra of the INP
indicate the presence of large beta-sheet folds. This may explain the tendency
of INPs to aggregate (Schmid, et al., 1997).

Bacterial ice-nucleation proteins are among the most active nucleants. The
three-dimensional theoretical model of such proteins, based on molecular
modeling, predicts a largely planar extended molecule, with one side serving
as a template for ice-like spatial orientation of water molecules and the
other side interacting with the membrane (Kajava and Lindow, 1993). This model
predicts also that INPs can form big aggregates by interdigitation.

In contrast to INPs, AFPs do not display any tendency to aggregation at
physiological concentrations. Moreover, the hydroxyl groups on the surface of
AFPs, combined with hydrophobic properties looks to be important for their
inhibition of the process of ice formation.\medskip

\begin{center}
\textbf{5.2. X-ray and other experimental data on AFP}
\end{center}

\textbf{\medskip}

The X-ray analysis with resolution of 1.25 \AA\ for crystals of globular type
III AFP with molecular mass of 7 kD was reported by Jia et al. (1996). It
reveals a remarkably flat aphipathic ice-binding site where five
hydrogen-bonding atoms match two ranks of oxygens on the [1010] ice prism
plane in the
$<$%
0001%
$>$%
direction, giving high ice-binding affinity and specificity. It looks that
type III AFP occupies a niche in the ice surface in which it covers the basal
plane while binding to the prism face.

The ice-binding mechanism of the long linear alpha-helical type I AFP (33 kD
as been attributed to their regularly spaced polar residues matching into the
ice lattice along a pyramidal plane (Sicheri and Yang, 1995).

The AFP of type III was linked through its N-terminals to thiodoxin (12 kD) or
maltose-binding protein (42 kD). The resultant 20-kD and 50-kD fusion proteins
were larger in diameter than free AFP and thus precluded any extensive [AFP -
AFP] contacts on the ice surface. Both fusion proteins had the same activity
as free AFP (III) at different concentration tested. This points out that AFP
do not require specific intermolecular interaction to bind to ice in a way,
inhibiting crystal growth (DeLuca et al., 1998).

AFP are structurally diverse molecules that share an ability to bind to ice
crystals and inhibit their growth. The type II fish AFP of Atlantic herring
and smelt are unique among known AFPs in their requirement of a cofactor for
antifreeze activity. These AFPs are homologous with the
carbohydrate-recognition domains of Ca2+-dependent (C-type) lectins and
require Ca2+ for their activity. To investigate the role of metal ions in the
structure and function of type II AFP, binding of Ca2+ and other divalent
cations to herring AFP was investigated (Ewart et al., 1996). The studies
showed that AFP has a single Ca2+-binding site.

Proteolytic protection studies and measurement of antifreeze activity revealed
a conformational change from a protease-sensitive and inactive apoAFP to a
protease-resistant active AFP upon Ca2+ binding. Other divalent metal ions
including Mn2+, Ba2+, and Zn2+ bind at the Ca2+-binding site and induce a
similar change.

An increase in tryptophan emission intensity at 340 nm also occurred upon Ca2+
addition. Whereas antifreeze activity appeared normal when Ca2+ or Mn2+ were
bound, it was much lower in the presence of other metal ions. When Ba2+ was
bound to the AFP, ice crystals showed a distinct difference in morphology.
These studies demonstrate that herring AFP specifically binds Ca2+ and,
consequently, adopts a conformation that is essential for its ice-binding activity.

The ability to control extracellular ice formation during freezing is critical
to the survival of freezing-tolerant plants. AFPs, having the ability to
retard ice crystal growth, were recently identified as the most abundant
apoplastic proteins in cold-acclimated winter rye (\emph{Secale cereale L.}) leaves.

The thermodynamics of intracellular ice nucleation are important in
low-temperature biology for understanding and controlling the process of cell
destruction by freezing. A new apparatus and technique for studying the
physics of intracellular ice nucleation was developed (Tatsutani, Rubinsky,
1998). Employing the principle of directional solidification in conjunction
with light microscopy, one can generate information on the temperature at
which ice nucleates intracellularly as a function of the thermal history which
the cells are encountered. The methods were introduced, and results with
primary prostatic cancer cells were described.

AFPs depress the freezing temperature of a solution in a non-colligative
manner, by arresting the growth of ice crystals. The kinetics of this effect,
as studied using a new technique called temperature gradient thermometry, are
consistent with an adsorption-mediated inhibitory mechanism (Chapsky,
Rubinsky, 1997). The results provided a new experimental basis for
understanding AFP interactions with ice.

A new microsensor that can analyze microliter volume samples was used to
measure the viscosity of aqueous solutions of antifreeze glycoproteins as a
function of temperature and concentration (Eto, Rubinsky, 1993). The results
showed that at physiological concentrations which naturally occur in fish,
AFPs significantly increased viscosity of aqueous solution. The probability
for ice nucleation is inversely proportional to viscosity. Therefore, the
increased viscosity could explain, in part, reports on the beneficial effects
of antifreeze glycoproteins during cryopreservation by vitrification. Reducing
the probability for ice nucleation could be also beneficial for the survival
of cold-water fish in their natural habitat. Millimolar concentrations of AFPs
increase viscosity of aqueous solution to values comparable with those of
conventional cryoprotectants in molar concentrations.

\bigskip

\begin{center}
\textbf{5.3 Molecular dynamics approach to the mechanism of AFP activity and
its biomimetics\medskip}
\end{center}

Computer assisted approach to design of biomimetic inhibitors controlling
crystal growth was used by Wierzbicki and Madura from University of South
Alabama, USA. Other efforts alike are the new drug development and the
entirely new methods of organ cryopreservation. The focus was on biomimetics
that modify crystal growth via stereospecific crystal surface inhibition.

Research efforts since the late 1960s have determined that the antifreeze
activity occurs through a unique non-colligative mechanism (DeVries et
al.,1984; Feeney et al., 1986). This non-colligative activity is evidenced by
a lowering of the freezing point without affecting the osmotic pressure (i.e.
water activity) or the melting temperature (Feeney et al.,1973; Tomchaney et
al., 1982). Although the antifreeze polypeptides do not function through a
colligative mode for dissolved solutes, they must act in some specific way to
alter the physical properties and interactions of the ice-water system.
Several possible mechanistic explanations for this observed phenomenon of
antifreeze activity may involve the solution phase itself, formation of solid
solutions, and/or the inhibition of ice-crystal growth at the ice solution
interface (Feeney et al., 1986). It is, however, this latter explanation that
has received the main attention. Accommodation of antifreeze molecules onto
the ice surface leads to perturbations in both the thermodynamics of the
surface interactions and the activation energy of the process. Ice propagation
becomes disrupted forcing ice growth to be limited to discrete areas lacking
the antifreeze molecules (Raymond and DeVries , 1972; 1977). This effect,
known as the Kelvin effect, has been recently accepted as an explanation of
the non-colligative freezing point depression mechanism of AFPs (Knight et
al., 1991).

Three different types of AFPs, namely of Type I, Type II, and Type III,
interacting stereospecifically with appropriate faces of ice crystal, induce
the modification of ice crystal shape and cessation of crystal growth within
well-defined temperature. For the alpha-helical Type I antifreeze protein, the
polypeptide-crystal interaction leads to hexagonal bipyramidal shape of ice crystal.

In a recent reviews by Feeney and Yeh (1993, 1996) several viable practical
applications of both native and/or synthetic antifreeze peptides were
discussed. These include such areas of application as recrystallization
inhibition, protection for non-polar fish, agricultural crops, as well as many
foodstuffs that require reduced temperatures or freezing conditions for long
term storage.

Since ice crystal growth is responsible for cellular/tissue damages within
biological materials, the inhibitors or modifiers of ice crystal growth,
analogous to the functional behavior of native antifreeze polypeptides, need
to be developed for the reduction of this damage.

Notwithstanding the big efforts, directed to elucidation of AFP mechanisms,
they are still quite obscure. Especially poorly understood is the AFPs%
\'{}%
influence on vicinal water: 3-60 \AA\ from biopolymer surface and even more
distant from surface water medium. The main difficulty of this problem was the
absence of right theoretical approach to analysis of distant water-protein
interaction, which is directly interrelated with the absence of general theory
of water and ice, including the mechanism of 1st order phase transition.

\textbf{It is evident from the literature data presented, that the mechanism
of action of AFPs cannot be understood by existing methods. The innovative,
principally new approaches must be applied for solving this problem}. \medskip

The advantage of our approach is that one of us (A. K.) during last 12 years
has elaborated the new Hierarchic Theory of Condensed Matter, general for
liquids and solids, published in a few books. This theory was verified
quantitatively on examples of water ice, using theory-based computer program:
''Comprehensive Analyzer of Matter Properties (CAMP)'' [copyright, 1997,
Kaivarainen] as it was partially illustrated at Sections 1- 4 of this article.
The full description of theory is located at the Archives of Los-Alamos: http://arXiv.org/abs/physics/01022086.

It will be illustrated below, how our Hierarchic theory can be applied to
explanation of mechanism of AFP action.

\textbf{\bigskip}

\begin{center}
\textbf{VI. HIERARCHIC APPROACH TO THEORY OF SOLUTIONS AND COLLOID
SYSTEMS\medskip}
\end{center}

The action of the dissolved molecules can lead to the shift of equilibrium to
the right or to the left for effectons (tr and lb) of solvent. In the former
case the lifetime of unstable state for primary effectons increases, and in
the latter case the stabilization of molecular associates (clusters) takes
place. The same is true for convertons%
\'{}%
equilibrium: [lb%
$\backslash$%
tr], reflecting [association%
$\backslash$%
dissociation] of water clusters (primary librational effectons). If the wave B
length of the dissolved molecule or atoms exceeds the dimensions of primary
effectons, then it must increase the degree of liquid association due to
MESOSCOPIC molecular Bose condensation. In the opposite case the ordering of
liquid structure decreases. \medskip

\textbf{In host-guest systems a following situations are possible:\smallskip\ }

1) guest molecules stabilize (a)-states of host \textbf{effectons} ( tr and
lb) and increases their dimensions.

As a result, the (a $\rightleftharpoons$ b) equilibrium of the
\textbf{effectons} and [lb $\rightleftharpoons$ tr] equilibrium of
\textbf{convertons }(see section II) becomes shifted leftward decreasing
potential energy of a system, corresponding to its stabilization effect;

2) guest molecules destabilize (a)-states of host effectons. The (a - b) and
[lb - tr] equilibrium of the primary effectons and convertons correspondingly
are shifted rightward, inducing general destabilization effect of the system;

3) guest and host molecules form separate individual effectons (mesophase)
without separation in two macrophases.\medskip

\textbf{The interaction of a solute (guest) molecule with librational solvent
effectons can be subdivided into two cases: when the rotational correlation
time of a guest molecule is less (A) and more (B) than the rotational
correlation time of librational effectons-coherent water clusters}. This time
is proportional to the mass and volume of the [lb] effecton. The number of
molecules in a librational primary effecton, depending on temperature: in
water it decreases from 280 till to 3 in the temperature interval 0-100
degrees of C (Fig.2). This means that the mass of primary [lb] effecton of
liquid water near freezing point is about 5000 D.

When the condition \textbf{(A)} is realized, small and neutral guest molecules
affect presumably only the translational effectons. In the second case
\textbf{(B)} the guest macromolecules can change the properties of both types
of effectons: translational and librational and shift the equilibrium [lb%
$\backslash$%
tr] of convertons to the left, stimulating the cluster formation. In
accordance with our model, the hydrophilic interaction is related to the shift
of the (a - b) equilibrium of translational effectons to the left. As far the
potential energy of the (a) state is less than that in the (b) state (Table
1), it means that such solvent-solute (host- guest) interaction will decrease
the potential and free energy of the solution. Hydrophilic interaction does
not need the realization of condition (II).

Hydrophobic interaction can be explained by the shift of the (a
$\rightleftharpoons$ b) equilibrium of librational effectons to the right.
Such a shift results in the increased potential energy of the system. The
dimensions of coherent water clusters, representing librational effectons
under condition (II) may even increase. However, the decrease in of entropy
(S) in this case is more than that in enthalpy (H) and, consequently, free
energy will increase:
\begin{equation}
\Delta G=\Delta H-T\Delta S\,>\,0 \tag{6.1}%
\end{equation}
This is a source of hydrophobic interaction, leading to aggregation of
hydrophobic particles.\smallskip

\begin{center}
\textbf{6.1. The new} \textbf{Clusterphilic Interaction of water with
macromolecules\smallskip}
\end{center}

A new kind of \textbf{Clusterphilic Interaction} was introduced (Kaivarainen,
1985, 1995; 2000) to describe the cooperative water cluster interaction with
nonpolar protein cavities. This idea has got support in the framework of our
hierarchic theory.

Clusterphilic interactions are related to:

1) the leftward shift of (a $\rightleftharpoons$ b) equilibrium of primary
librational effectons under condition (B);

2) to the similar shift of the equilibrium of [lb $\rightleftharpoons$ tr] of
convertons; and

3) increasing of primary [lb] effectons dimensions due to water molecules
immobilization near the surface of macromolecules.

The latter effect is a result of decreasing of the rotational correlation time
of librational effectons and decreasing of the most probable impulses of water
molecules, related to librations under the effect of big enough guest particles.\medskip

\begin{center}
\textbf{Clusterphilic interactions can be subdivided into:\smallskip}
\end{center}

1. Intramolecular - when water cluster is localized in the ''open'' states of
big interdomain or intersubunit cavities;

2. Intermolecular clusterphilic interactions. Intermolecular clusterphilic
interactions can be induced by very different sufficiently big macromolecules. \smallskip

Clusterphilic interactions can play an important role in the self-organization
of biosystems, especially multiglobular allosteric enzymes, microtubules and
the actin filaments. Cooperative properties of the cytoplasm, formation of
thixotropic structures, signal transmission in biopolymers, membranes and
distant interactions between different macromolecules can be mediated by both
types of clusterphilic interactions.

Clusterphilic interactions and possible self-organization in colloid systems -
thixotropic structure formation, are promoted mainly by decreasing of
contribution to potential energy, related to librations in the presence of
macromolecules. Hydrophilic interactions are the result of decreasing of the
contribution of potential energy related to translations of water molecules.
Hydrophobic interaction is a consequence of increasing of potential energy of
system, related to rightward [a $\rightleftharpoons$ b] equilibrium shift of
both kind of the effectons: librational and translational under the influence
of guest molecules, as far $V_{a}<V_{b}$. Macromolecules or polymers with
molecular mass less than 2 kD do not satisfy the condition (B) at the ambient
and cannot stimulate the growth of librational effectons.\medskip

\begin{center}
\textbf{6.2. The multi-fractional nature and properties of interfacial
water\smallskip}
\end{center}

We can present here a classification and description of FOUR interfacial water
fraction properties, based on the hierarchic model:

1. First fraction - primary hydration shell with maximum energy of interaction
with surface.

The structure and dynamics of this 1st fraction can differ strongly from those
of bulk water. Its thickness (3-10 \AA) corresponds to 1-3 solvent molecules;

2. Second fraction - vicinal water (VW) is formed by elongated primary lb
effectons with saturated hydrogen bonds. It is a result of lb effecton
adsorption on the primary hydration shell from the bulk volume. The thickness
of this fraction of interfacial water: (30-75 \AA) corresponds to 10-25
molecules and is dependent on the temperature, dimensions of colloid particles
and their surface mobility. VW can exist in rigid pores of corresponding dimensions;

3. Third fraction of interfacial water - the surface-stimulated
Bose-condensate (SSBC), represented by 3D network of primary librational
effectons has a thickness of (50-300 \AA\ ). It is a next hierarchical level
of interfacial water self-organization on the surface of second fraction (VW).
The time of gradual formation of this 3D net of linked to each other coherent
clusters (strings of polyeffectons), is much longer than that of VW and it is
more sensitive to temperature and other perturbations. The second and third
fractions of interfacial water can play an important role activity of
biological cells activity;

4. The biggest and most fragile fourth fraction of interfacial water can be a
result of slow orchestration of bulk primary effectons in the volume of
primary (electromagnetic) lb deformons. The primary deformons appears as a
result of three standing IR photons (lb) interception. Corresponding IR
photons are superradiated by the enlarged lb effectons of vicinal water. The
linear dimension of primary IR deformons is about half of librational IR
photons, i.e. 5 microns (5*10\symbol{94}4 \AA). This ''superradiation
orchestrated water (SOW)'' fraction easily can be destroyed not only by
temperature increasing, ultrasound and Brownian movement, but even by
mechanical shaking. The time of spontaneous reassembly of this fraction after
destruction has an order of hours and is dependent strongly on temperature,
viscosity and dimensions of colloid particles. The processes of
self-organization of third (SSBC) and forth (SOW) fractions can be
interrelated by feedback interaction. \textbf{\medskip}

\begin{center}
\textbf{VII. Theoretical model of the antifreeze proteins (AFP) action\medskip}
\end{center}

Hierarchic model of the formation of \ ''clustron'' structures by AFP and
their possible influence on water and ice-formation. We will proceed from our
model of 1st order phase transition, described above and experimental data
available. The preliminary model, we are going to elaborate should be able to
explain the following results:

a) decreasing of freezing point of water at relatively small concentrations of
purified AFP of about (0.5-1) mg/ml at weak buffers at physiological pH (7.0 -
7.5). At these concentrations the water activity and osmotic pressure
practically do not change thus explaining the hysteresis between freezing and melting;

b) changing of shape of ice crystals from hexagonal, pertinent for pure water,
to elongated bipyramidal one;

c) accumulation of AFPs at the interface between ice and water;

d) ''controversy'' of AFPs%
\'{}%
surface properties, i.e. the presence of regular hydroxyl groups necessary for
hydrogen bonding, like threonine residues and their ability to inhibit the ice
microcrystal growth;

e) significant increase of viscosity of water in presence of AFPs, directly
correlated with their antifreezing activity.\medskip

\textbf{Our mechanism of AFP action, is related to specific influence of these
proteins on the properties of primary librational effectons and translational
mobility of water molecules. }

\textbf{Our model includes the following stages: }

\textbf{1.} Tendency of AFP to surround the primary [lb] effectons with their
four-coordinated ideal ice structure, resulted from certain stereo-chemical
complementary between distribution of side group on the surface of AFP and ice;

\textbf{2.} Increasing the dimensions of this coherent clusters due to water
molecules librations immobilization near the surface and vicinal water,
corresponding decreasing of the most probable [lb] impulse (momentum) and
increasing of de Broglie wave (wave B) length. Such enlarged water clusters
(primary [lb] effecton) surrounded by macromolecules we named
''\textbf{clustron}'' (Kaivarainen 1995, 1998);

\textbf{3.} Changing the symmetry of librational water molecules, thermal
oscillations, from almost isotropic to anisotropic ones. Thus it means
redistribution of thermal energy between three librational degrees of freedom
again due to special spatial/dynamic properties of AFP surface. The
consequence of this effect can be the change of ice-nucleation centers shape
from hexagonal to bipyramidal, making the process of freezing less favorable.
It should result in corresponding change in ice microcrystal form;

\textbf{4.} Bordering of the enlarged primary [lb] effectons by AFP provides
the ''insulation effect'', preventing the association of these effectons to
big enough nucleation centers, which normally accompany the water-ice
transition, judging from our computer simulations;

\textbf{5}. Increasing the number of defects in the process of ice lattice
formation in presence of AFP, as far the freezing temperature of water inside
clustron can be much lower than in free bulk water. It is known experimental
fact that the freezing point of water in pores or cavities with dimensions of
about nanometers is much lower, than in bulk water (Kaivarainen, 1985). Such
the interface effect may explain accumulation of AFP between solid and liquid phase;

\textbf{6.} Shifting the equilibrium between the ground - acoustic (a) state
of enlarged primary [lb] effectons and optic (b) state to the latter one. This
shift increases the potential energy of the affected [lb] effecton because the
potential energy of (b) state is higher that of (a) and the kinetic energy of
both states are the same in accordance to our model;

\textbf{7}. Increasing the probability and frequency of excitation of [lb-tr]
convertons, i.e. [association - dissociation] of water molecules in the volume
of \textbf{clustron}, accompanied by water density fluctuations and
corresponding pulsation of the clustron's volume;

\textbf{8.} Radiation of the ultrasound (US) waves by \textbf{clustrons,}
pulsing with frequency of convertons (10$^{6}$-10$^{7}$) s$^{-1}$. These US
waves destroy the ice-nucleation centers and activate the translational
mobility (impulse) of water molecules, decreasing the translational wave B
length $(\lambda_{tr}=h/mv_{tr})$. In accordance to condition of the 1st order
phase transition (4.5) this effect decreases the freezing point of water. The
melting point remains unchanged, explaining the T-hysteresis provided by AFP,
because the clustron pulsation in solid phase of system are inhibited;

\textbf{9.} Tendency to association of pulsing clustrons to form
thixotropic-like structures in liquid phase, could be a consequence of
Bjorkness hydrodynamic attractive forces, existing between pulsing particles,
which radiate the acoustic density waves and dipole-dipole interaction between
clustrons. The formation of thixotropic structures, resulting from clustron
association, explains the increasing of viscosity in AFP solutions.\medskip

This preliminary model of AFP action, based on hierarchic theory (Kaivarainen,
2001), includes the mechanism of interfacial water formation and theory of
solutions, described above. The goal of our future experimental work is to
confirm the main idea of proposed mechanism and evaluate the role of all
contributions, listed above.

\textbf{The self-organization of thixotropic- like structures in aqueous
solutions of macromolecules} was experimentally shown by Giordano et al.,
(1981). It seems to be very general phenomena. The results obtained from
viscosity, acoustic and light-scattering measurements showed the existence of
long-range structures that exhibit a thixotropic behavior. This was shown for
solutions of lysozyme, bovine serum albumin (BSA), hemoglobin and DNA. Ordered
structure builds up gradually in the course of time to become fully developed
after more than 10-15 hours. When a sample is mechanically shaken this type of
self-organization is destroyed. The ''preferred distance'' between
macromolecules in such an ordered system is about 50 \AA\ as revealed by small
angle neutron scattering (Giordano et al., 1991). It is important that this
distance can be much less than the average statistical distance between
proteins at low molar concentrations.\medskip

\begin{center}
\textbf{7.1. } \textbf{Consequences and predictions of proposed model of AFP
action\medskip}
\end{center}

1. The radius of clustrons enlarged primary [lb] effectons, surrounded by AFP
can be about 30-50 \AA, depending on properties of surface (geometry,
polarity), temperature, pressure and presence the perturbing solvent
structural agents.

2. Water, involved in clustron formation should differ by number of physical
parameters from the bulk water. It should be characterized by:

a) lower density;

b) bigger heat capacity;

c) bigger sound velocity

d) bigger viscosity;

e) smaller dielectric relaxation frequency, etc.\smallskip

The formation of thixotropic structure in AFP-water systems should be
accompanied by non-monotonic spatial distribution of AFP in the volume due to
interaction between clustrons.

The compressibility of primary [lb] effectons should be lower and sound
velocity higher than that of bulk water. It is confirmed by results of
Teixeira et. al. (1985), obtained by coherent- inelastic-neutron scattering.
This was proved in heavy water at 25 C with solid-like collective excitations
with bigger sound velocity than in bulk water. These experimental data can be
considered as a direct confirmation of primary librational effectons existence.

The largest decrease of water density occurs in pores, containing enlarged
primary librational effectons, due to stronger immobilization of water
molecules. The existing experimental data point, that the freezing of water
inside pores occurs much below 0 $^{0}$C.

\textbf{We can conclude, that most of consequences of proposed model of AFP
action is in a good accordance with available experimental data and the
verification of the rest ones is a matter of future work. }

\bigskip

\textbf{\medskip References:}

\begin{quotation}
Aksnes G., Libnau O. $\left(  1991\right)  \,$ Acta Chem.Scand.$,\,$%
45,\thinspace463.

Ashkroft N., Mermin N. (1976). Solid state physics. N.Y. (Helt, Rinehart and Winston).

Babloyantz A. (1986), Molecules, Dynamics and Life. An introduction to
self-organization of matter. John Wiley \& Sons, Inc. New York, pp.320.\ \ 

Chapsky L, Rubinsky B., Kinetics of antifreeze protein-induced ice growth inhibition.

FEBS Lett 1997, 412:1, 241-244.

Chernikov F.R. $\left(  1990\right)  $, Biofizika (Russia$),\;$35,\ 711.

Clegg J.S. and Drost-Hansen W. $\left(  1991\right)  \;$Elsevier Science Publ.
vol.1, Ch.1, pp.1-23.

DeLuca C.I., Comley R., Davies P.L. Antifreeze proteins bind independently to
ice. Biophys. J. 1998, Mar 74:3, 1502-1508.

Del Giudice E., Doglia S., Milani M. and Vitello G. (1983) Nucl. Phys. B275
[Fs 17], 185.

Del Giudice E., Doglia S., Milani M. (1988), Physica Scripta. 38, 505.

Del Guidice E., Preparata G., Vitello G. $\left(  \text{1989}\right)  \;$
Phys. Rev. Lett. $,61,$1085.

DeVries, A.L. Role of glycopeptides and peptides in inhibition of
crystallization of water in polar fishes. Philos.Trans. R. Soc. Lond. B 304,
575-588. 1984.

Dicke R.H. $\left(  1954\right)  \;$ Phys.Rev. 93, 99-110$.$

Drost-Hansen W. In: Colloid and Interface Science. Ed. Kerker M. (Academic
Press, New York, 1976), p.267-280.

Duman J.G, Li N, Verleye D, Goetz F.W, Wu D.W, Andorfer C.A, Benjamin T,

Parmelee D.C. Molecular characterization and sequencing of antifreeze proteins
from larvae of the beetle Dendroides canadensis. J. Comp. Physiol. [B] 1998,
Apr 168:3 225-232.

Eisenberg D., Kauzmann W. The structure and properties of water. Oxford
University Press, Oxford, 1969.

Etzler F.M., Conners J.J. $\left(  1991\right)  \;$Langmuir , 7, 2293.

Etzler F.M., White P.J. J. $\left(  1987\right)  \;$Colloid and Interface
Sci., 120, 94.

Eto T.K, Rubinsky B. Antifreeze glycoproteins increase solution viscosity.

Ewart K.V, Yang D.S. , Ananthanarayanan V.S, Fletcher G.L, Hew C.L.

Ca2+-dependent antifreeze proteins. Modulation of conformation and activity by
divalent metal ions. J Biol Chem 1996, v. 271, N:28, 16627-16632.

Feeney, R.E., T.S. Burcham, and Y.Yeh. Antifreeze glycopeptides from polar
fish blood. Ann. Rev. Biophys. Chem. 15, 59-78. 1986.

Feeney, R.E. and R. Huffman. Depression of freezing point by glycoprotein from
an Ant-arctic fish. Nature 243, 357-359. 1973.

Feeney, R.E. and Y.Yeh. Antifreeze proteins: properties, mechanism of action,
possible applications. Food Technology, 82-89, 1993.

Fine R.A., Millero F.J. $\left(  1973\right)  \,$ J.Chem.Phys. , 59, 5529$.$

Frontas'ev V.P., Schreiber L.S. (1966) J. Struct. Chem. (USSR$)\,\;$6, 512$.$

Fr\"{o}hlich H. (ed.)\ (1988)\ Biological coherence and response to external
stimuli. Springer, Berlin.

Grawford F.S. (1973) Waves. Berkley Physics Course. Vol.3. McGraw- Hill Book
Co., N.Y.

Haken H. (1990), Synergetics, computers and cognition. Springer, Berlin.

Haymet A.D. et al., 1998, FEBS Letters, , 1998, 430:301-306, July 3.

Hon W.C, Griffith M, Mlynarz A, Kwok Y.C, Yang D.S.,

Antifreeze proteins in winter rye are similar to pathogenesis-related proteins.

Plant Physiol. 1995, Nov, 109:3 879-889.

Jia Z., DeLuca CI., Chao H., Davies P.L. Structural basis for the binding of
globular antifreeze protein to ice. Nature 1996, Nov.21, v. 384:6606, 285-288.

Johri G.K., Roberts J.A. $\left(  1990\right)  \;$J. Phys. Chem. $,94,\,7386.$

Kaivarainen A. (2001) Series of articles at the Archives of Los-Alamos: http://arXiv.org/find/physics/1/au:+Kaivarainen\_A/0/1/0/all

Kaivarainen A. (1995) Book: Hierarchic concept of matter anf field. Water,
biosystems and elementary particles. New York, NY, ISBN 0-9642557-0-7.

K\"{a}iv\"{a}r\"{a}inen A. (1996). Paper in Proc.of 2nd. Ann Adv. Water Sci.
Symp. October 4-6, Dallas, Texas.

K\"{a}iv\"{a}r\"{a}inen A., Fradkova L., Korpela T. $\left(  1993\right)  ,$
Acta Chem.Scand. $47,\,456.$

Kaiv\"{a}r\"{a}inen A. $\left(  1989\right)  ,$\ J. Mol. Liq. $\,41,\,53$.

Kaivarainen A.I. (1985) Solvent-dependent flexibility of proteins and
principles of their function. D.Reidel Publ.Co., Dordrecht, Boston, Lancaster,
290 p.

Kell G.S. $\left(  1975\right)  ,$ J.Chem.Eng. Data \ 20, 97.

Kikoin I.K. (Ed.) (1976), Tables of physical values. Atomizdat, Moscow.

Knight, C. A., C. C. Cheng. and A. L. DeVries. 1991. Adsorption of a-helical
antifreeze peptides on specific ice crystal surface planes. Biophysical
Journal. 59, 409- 418.

Li N., Andorfer Cathy A. and Duman J. G. Enhancement of insect antifreeze
protein activity by solutes of low molecular mass. J. Exp. Biol. 1998, v.201,
15, 2243-2251.

Prokhorov A.M. (Ed) (1988). Physical encyclopedia. Soviet encyclopedia. Moscow.

Raymond, J.A. and A.L. DeVries. Cryobiology 9, 541-547. 1972.

Raymond, J.A. and A.L. DeVries. Adsorption inhibition as a mechanism of
freezing resistance in polar fishes. Proc. Nat. Acad. Sci. USA. 74, 2589-2593. 1977.

Roberts J. and Wang F. J. $\left(  1993\right)  ,\,$Microwave Power and
Electromagnetic Energy, $28,196$.

Roberts J., Zhang X. \& Zheng Y. J. (1994), J. Chem. Phys. Determination of
hydrogen bonding in water clusters through conductivity (d.c.) measurements of
aqueous solutions of NaCl.

Sicheri F. and Yang D.S., Ice-binding structure and mechanism of an antifreeze
protein from winter flounder, Nature 1995, v.375, N 6530, 427-431.

Tatsutani K., Rubinsky B. A method to study intracellular ice nucleation.

J Biomech Eng 1998, 20:1, 27-31.

Tomchaney, A.P., J.P. Morris, S.H. Kang, and J.G. Duman. Purification,
composition and physical properties of a thermal hysteresis of antifreeze
protein from larvae of the beetle, Tenebrio molitor, Biochemistry 21, 716-721. 1982.

Yeh, Y. and R.E Feeney. Antifreeze Proteins: Structures and Mechanisms of
Function. Chemical Reviews, 96, 601-617, 1996.

Umezava H., Matsumoto H. and Tachiki L. (1982). Thermo-field dynamics and
condensate states (North-Holland, Amsterdam).

Umezawa H. (1993). Advanced Field Theory: Micro, Macro and Thermal Physics.
American Institute of Physics, New York.

Wang H., Lu B. \& Roberts J.A. (1994) Molecular materials.

Watterson J. (1988), Mol.Cell.Biochem. 79, 101-105.

Watterson J.\ (1988), BioSystems 22, 51.
\end{quotation}
\end{document}